\documentclass[aps,nofootinbib,twocolumn,preprintnumbers,superscriptaddress]{revtex4-2}

\usepackage[T1]{fontenc} 
\usepackage[utf8]{inputenc}

\usepackage[dvipdfm]{graphicx}
\usepackage{amssymb}
\usepackage{xcolor}
\usepackage{amsmath}
\usepackage{slashed}
\usepackage{hyperref}
\hypersetup{
    unicode=true,
    colorlinks=true,
    linkcolor=blue,
    citecolor=red,
    filecolor=black,
    urlcolor=blue
    }

\begin{document}\sloppy

\vspace*{1mm}

\title{
Boltzmann or Bogoliubov?\\
Approaches Compared in Gravitational Particle Production
}

\author{Kunio Kaneta}
\email{kkaneta@lab.twcu.ac.jp}
\affiliation{Department of Mathematics, Tokyo Woman’s Christian University, Tokyo 167-8585, Japan}
\author{Sung Mook Lee}
\email{sungmook.lee@yonsei.ac.kr}
\affiliation{Department of Physics \& IPAP \& Lab for Dark Universe, Yonsei University, Seoul 03722, Korea}
\author{Kin-ya Oda}
\email{odakin@lab.twcu.ac.jp}
\affiliation{Department of Mathematics, Tokyo Woman’s Christian University, Tokyo 167-8585, Japan}

\vspace{0.5cm}

\date{\today}

\begin{abstract} 
Gravitational particle production is a minimal contribution to reheating the Universe after the end of inflation.
To study this production channel, two different approaches have commonly been considered, one of which is based on the Boltzmann equation, and the other is based on the Bogoliubov transformation.
Each of these has pros and cons in practice. 
The collision term in the Boltzmann equation can be computed based on quantum field theory in the Minkowski spacetime, and thus many techniques have been developed so far.
On the other hand, the Bogoliubov approach may deal with the particle production beyond the perturbation theory and is able to take into account the effect of the curved spacetime, whereas in many cases one should rely on numerical methods, such as lattice computation.
We show by explicit numerical and analytical computations of the purely gravitational production of a scalar that these two approaches give consistent results for particle production with large momenta during reheating, whereas the Boltzmann approach is not capable of computing particle production out of vacuum during inflation.
We also provide analytic approximations of the spectrum of produced scalar with/without mass for the low momentum regime obtained from the Bogoliubov approach.

\end{abstract}

\maketitle

\setcounter{equation}{0}

\section{Introduction}

Reheating is the necessary phase that should be realized in any inflationary models of modern cosmology \cite{Olive:1989nu,Linde:1990flp,Lyth:1998xn,Linde:2000kn,Martin:2013tda,Martin:2013nzq,Martin:2015dha}.
After the exponential expansion of space, the Universe is fulfilled by radiation, following the reheating era during which particles of the Standard Model are produced.
The thermal plasma of the produced particles sets the stage of Big Bang Nucleosynthesis by requiring that the reheating temperature, $T_{\rm RH}$, is greater than $\sim$ 1 MeV \cite{Alpher:1948ve,Walker:1991ap,Olive:1999ij,Fields:2006bzp,Fields:2014uja,Steigman:2007xt,Cyburt:2001pp,Nollett:2000fh,Burles:2000zk,Vangioni-Flam:2000yfu,Cyburt:2001pq,Cyburt:2003fe,Coc:2003ce,Cuoco:2003cu,Serpico:2004gx,Cyburt:2004cq,Descouvemont:2004cw,Iocco:2008va,Pisanti:2007hk,Coc:2011az,Cyburt:2008kw,Coc:2014oia,Coc:2015bhi,Pitrou:2018cgg,Cyburt:2015mya,Fields:2019pfx,Yeh:2020mgl}.
If one supposes that the baryon asymmetry of the Universe is generated through lepton asymmetry \cite{Fukugita:1986hr,Barbieri:1999ma,Davidson:2002qv,Pilaftsis:2003gt,Raidal:2004vt,Nardi:2006fx,Abada:2006ea,Buchmuller:2012wn}, $T_{\rm RH}$ needs to be greater than the critical temperature of the electroweak phase transition \cite{Khlebnikov:1988sr,Harvey:1990qw}.

For the Universe to be dominated by radiation after the reheating completes, a sufficient amount of energy stored in the inflaton sector should be transferred into the thermal bath.
The energy transfer is most commonly realized in perturbative inflaton decay through a direct coupling between inflaton and matter particles. \footnote{However, see, for instance, Refs. \cite{Kofman:1997yn,Bassett:1998wg,Felder:1998vq,Greene:1998nh,Bassett:1999mt,Felder:1999pv,Chung:1999ve,Greene:2000ew,Peloso:2000hy,Dufaux:2006ee,Frolov:2010sz,Jedamzik:2010dq,Amin:2014eta,Giblin:2019nuv,Fan:2021otj,Garcia:2021iag} for non-perturbative contribution in the energy transfer, the process known as preheating.}
In general, the decay of inflaton does not occur instantaneously \cite{Giudice:2000ex,Chung:1998rq,Dudas:2017rpa,Garcia:2017tuj,Chen:2017kvz}.
Such non-instantaneous inflaton decay leads to non-trivial evolution of the temperature and the equation of state until reaching the radiation-dominated era.
Moreover, the time evolution of the inflaton energy density strongly depends on the shape of the inflaton potential during the reheating \cite{Garcia:2020eof,Garcia:2020wiy} as well.
The thermalization of the produced particles is often assumed to be instantaneous in the literature, whereas it is not always the case that the thermalization is achieved at once after particle production happens from, for instance, the inflaton decay \cite{Davidson:2000er,Harigaya:2013vwa,Harigaya:2014waa,Mukaida:2015ria,Garcia:2018wtq,Harigaya:2019tzu}.

Both reheating and subsequent thermalization may have direct impact on new physics beyond the Standard Model such as generating dark matter particles, especially a so-called freeze-in massive particle (FIMP) dark matter \cite{Hall:2009bx,Chu:2011be,Mambrini:2013iaa,Chu:2013jja,Kaneta:2016vkq,Kaneta:2016wvf,Kaneta:2017wfh,Bernal:2017kxu,Biswas:2018aib,Bernal:2019mhf,Kaneta:2019zgw,Bernal:2020qyu,Bernal:2020gzm,Anastasopoulos:2020gbu,Brax:2020gqg,Brax:2021gpe,Kaneta:2021pyx,Ghosh:2022hen}.
Among various portal couplings to produce FIMP out of thermal bath, the graviton exchange is the minimal contribution \cite{Garny:2015sjg,Garny:2017kha,Tang:2017hvq,Chianese:2020yjo,Chianese:2020khl,Redi:2020ffc,Bernal:2018qlk}, namely, annihilation of the thermal bath particles to produce a pair of dark matter through a single graviton exchange. However, this contribution turns out to be subdominant, compared to the inflaton annihilation, instead of the thermal bath particles, to a pair of dark matter by exchanging a graviton~\cite{Chung:1998bt,Kolb:1998ki,Chung:2001cb,Ema:2018ucl,Chung:2018ayg,Ema:2019yrd,Ahmed:2020fhc,Gross:2020zam,Kolb:2020fwh,Mambrini:2021zpp,Ling:2021zlj,Haque:2021mab,Clery:2021bwz,Haque:2022kez,Clery:2022wib,Aoki:2022dzd}.

In this context, gravitational particle production has gained renewed attention since the first study of particle creation in curved spacetime has been done long ago \cite{Parker:1969au}.
Since then, particle production has been studied based on quantum field theory in curved spacetime by utilizing the Bogoliubov transformation, which we will call the "Bogoliubov approach" in the following.
So far, this is the most developed framework to keep track of the evolution of particle spectrum produced during inflation and reheating.
The same approach has been applied to compute the gravitational dark matter production from inflaton oscillation during the reheating~\cite{Chung:1998bt,Kolb:1998ki,Chung:2001cb,Ema:2018ucl,Chung:2018ayg,Ema:2019yrd,Ahmed:2020fhc,Gross:2020zam,Kolb:2020fwh}.

On the other hand, in the context of dark matter production, it is more conventional to compute particle production based on the Boltzmann equation with the collision terms originally obtained in the framework of quantum field theory in flat spacetime, which we will call the "Boltzmann approach". This is because in many circumstances dark matter production takes place when the Universe undergoes power expansion (such as matter-dominated or radiation-dominated epoch), and hence quantum field theory in Minkowski spacetime becomes a good approximation.
This framework has been used to compute the gravitational dark matter production \cite{Garny:2015sjg,Garny:2017kha,Tang:2017hvq,Chianese:2020yjo,Chianese:2020khl,Redi:2020ffc,Bernal:2018qlk,Mambrini:2021zpp,Clery:2021bwz,Clery:2022wib}.

The Bogoliubov and Boltzmann approaches are supposed to give identical results in certain cases of particle production, especially when dark matter is dominantly produced.
However, it is far from clear to what extent this claim is justified because these two approaches are based on different premises.
In the Bogoliubov approach, particle production takes place since the vacuum is in general time-dependent in curved spacetime.
By keeping track of the time evolution of the vacuum, one may observe the particle production out of the vacuum.
It is also important to note that it is not always the case that the particle production is analytically computable for a given spacetime geometry.

In contrast to the Bogoliubov approach, the vacuum being a constant of time is a prerequisite in the Boltzmann approach.
The particle production is instead described by scatterings of fields on Minkowskian geometry, and thus the analytic computation is much easier than the Bogoliubov approach in many cases. Therefore, at the first glance, there is no guarantee that these different approaches give the same physics result.

In this paper, we compare these two approaches by considering the gravitational production of a minimally coupled real scalar. We develop an analytic method to show that these approaches are certainly equivalent for the produced scalar particle whose momentum is greater than the inflaton mass. On the other hand, for smaller momentum ranges, the production rate of the particles in two formalisms are significantly different, because of large non-adiabacity of the regime.
Through further analytic computation of Bogoliubov approach, we derive the power-law spectrum of the particle produced out of the vacuum during reheating, which is the contribution beyond the Boltzmann approach.

The paper is organized as follows.
In section \ref{sec:setup} we explain the setup to discuss the gravitational production of a scalar field during the reheating, in the Boltzmann approach in section \ref{sec:Boltzman} and in the Bogoliubov approach in section \ref{sec:Bogoliubov_high-k}.
After discussing the vacuum contribution to gravitational particle production in section \ref{sec:Bogoliubov_low-k}, we summarize our results and address possible applications in section \ref{sec:summary}.

\section{Set-Up}
\label{sec:setup}
Before getting into the detail of the comparison between the Boltzmann and Bogoliubov approaches, we set a stage for what we will consider of gravitational particle production in each approach. Our framework consists of two parts: the Einstein-Hilbert gravity (spin-2) and a minimally coupled real scalars $\phi$ and $\chi$ (spin-0), i.e., $S=S_2 + S_0$ where
\begin{align}
    &
    S_2 = -\frac{M_P^2}{2}\int d^4x\sqrt{-g}R,\\
    &
    S_0 = \int d^4x\sqrt{-g}\left[{\cal L}_\phi + {\cal L}_\chi\right],\\
    &
    {\cal L}_\phi = \frac{1}{2}g^{\mu\nu}\partial_\mu\phi\partial_\nu\phi - V(\phi),\\
    &
    {\cal L}_\chi = \frac{1}{2}g^{\mu\nu}\partial_\mu\chi\partial_\nu\chi - \frac{1}{2}m_\chi^2\chi^2.
\end{align}
Here, we introduced the reduced Planck mass $M_P\equiv 1/\sqrt{8\pi G}\simeq 2.4\times 10^{18}~{\rm GeV}$ with $G$ the Newton constant, and $R$ is the Ricci scalar as a function of the Riemannian metric $g_{\mu\nu}$ whose determinant is denoted by $g\equiv {\rm det}(g_{\mu\nu})$.\footnote{Throughout the paper we work in the sign convention of mostly-minus metric, i.e., $\eta_{\mu\nu}={\rm diag}(+1,-1,-1,-1)$ for the Minkowskian metric $\eta_{\mu\nu}$.}

In the spin-0 sector, $\phi$ is the inflaton with potential $V(\phi)$, and $\chi$ is a spectator scalar field produced from the inflaton during the reheating. While our analytic results are quite independent to specific form of the potential, we take so-called T-model \cite{Kallosh:2013hoa} potential as an explicit example for numerical results:
\begin{align}
    V(\phi) &= 6\lambda M_P^4\tanh^2\left(\frac{\phi}{\sqrt{6}M_P}\right).
\end{align}
Note that after the end of inflation, $\phi$ oscillates about $\phi=0$ with a quadratic form of potential
\begin{align}
    V(\phi \ll M_P) \simeq \frac{1}{2}m_\phi^2\phi^2
\end{align}
with $m_\phi^2\equiv 2\lambda M_P^2$.
The potential coupling $\lambda$ is determined such that the amplitude of the curvature power spectrum $A_S$ is obtained at a number of e-folds $N$ through \cite{Garcia:2020wiy}
\begin{align}
    \lambda \simeq \frac{18\pi^2 A_S}{6N^2}.
\end{align}
In our analysis, we take $\ln(10^{10}A_S)=3.044$ for the pivot scale $k=0.05~{\rm Mpc}^{-1}$ \cite{Planck:2018jri} and assume $N=55$, yielding $\lambda = 2.1\times10^{-11}$ and $m_\phi\simeq 1.5\times10^{13}$ GeV.

\section{Particle production in the Boltzmann approach}
\label{sec:Boltzman}

The Boltzmann approach (when computing the collision term) assumes that the background geometry is Minkowskian, in which the gravitational production of $\chi$ takes place through $\phi\phi\to\chi\chi$ by exchanging a single graviton.
Note that this process exists during the reheating when $\phi$ coherently oscillates, but it disappears once the reheating completes as $\phi$ decays away.
It should be emphasized that the initial state $\phi$ is not a sigle particle, but a coherently oscillating Bose-Einstein condensate. Namely, $\phi$ does not have spatial momentum as if it sits at its rest frame.

To consider the gravitational particle production, we should start with defining the graviton field $h_{\mu\nu}$ by linearizing the gravity sector as $g_{\mu\nu}\simeq \eta_{\mu\nu} + 2h_{\mu\nu}/M_P$.
By introducing the stress-energy-momentum tensor $T_{\mu\nu} \equiv (2/\sqrt{-g})\delta S/\delta g^{\mu\nu}$, the action may be written as
\begin{align}
    &
    S\simeq 
    \int d^4x\left[ {\cal L}_{h,{\rm kin}} + {\cal L}_\phi + {\cal L}_\chi
    -\frac{1}{M_P}h^{\mu\nu}(T^\phi_{\mu\nu} + T^\chi_{\mu\nu})
    \right],
\end{align}
where the graviton kinetic term ${\cal L}_{h,{\rm kin}}$ is given by
\begin{align}
    &
    {\cal L}_{h,{\rm kin}} = -\frac{1}{2}h^{\mu\nu} P_{\mu\nu\alpha\beta}\Box h^{\alpha\beta},\\
    &
    P_{\mu\nu\alpha\beta} \equiv \frac{\eta_{\mu\alpha}\eta_{\nu\beta} + \eta_{\mu\beta}\eta_{\nu\alpha} - \eta_{\mu\nu}\eta_{\alpha\beta}}{2},
\end{align}
with $\Box\equiv \eta^{\mu\nu}\partial_\mu\partial_\nu$, leading to the graviton propagator for a momentum $q$: $\Pi^{\mu\nu\alpha\beta}(q)=P^{\mu\nu\alpha\beta}/q^2$.

The stress-energy-momentum tensor for scalar $\chi$ is given by
\begin{align}
    T^\chi_{\mu\nu} &=
    \partial_\mu\chi\partial_\nu\chi-\frac{1}{2}\eta_{\mu\nu}(\partial^\alpha\chi\partial_\alpha\chi-m_\chi^2\chi^2).
\end{align}
For $T^\phi_{\mu\nu}$, since $\phi$ is a coherently oscillating condensate, it is convenient to define the Fourier series by
\begin{align}
    T^\phi_{\mu\nu} &= \sum_{n=-\infty}^\infty T^\phi_{n,\mu\nu}e^{-in\omega t},
\end{align}
where
\begin{align}
    T^\phi_{n,\mu\nu} &\equiv
    2K_n u_\mu u_\nu - \eta_{\mu\nu}(K_n - V_n),\\
    u_\mu &\equiv (1,0,0,0)_\mu,
\end{align}
with $K_n$ and $V_n$ representing the Fourier coefficients for the kinetic and potential terms. Note that for $V=(1/2)m_\phi^2\phi^2$, we have $\omega=m_\phi$.

For the $n$-th mode of the $\phi$ oscillation, we may write the transition amplitude of $\phi\phi\to h_{\mu\nu}\to\chi(p_A)\chi(p_B)$, ${\cal M}_n$, as
\begin{align}
    {\cal M}_n &=
    \frac{1}{M_P^2}T^\phi_{n,\mu\nu} \Pi^{\mu\nu\alpha\beta}(n\omega)\nonumber\\
    &\qquad \times [p_{A\alpha}p_{B\beta}+p_{A\beta}p_{B\alpha}-\eta_{\alpha\beta}(p_A\cdot p_B + m_\chi^2)]\nonumber\\
    &=
    \frac{V_n}{M_P^2}\left(1+\frac{2m_\chi^2}{(n\omega)^2}\right).
    \label{eq:M_n}
\end{align}
Transition probability with the amplitude ${\cal M}_n$ may be defined by
\begin{align}
    dP^{(n)}_{\phi\phi\to AB} \equiv & \frac{d^3p_A}{(2\pi)^32p_A^0}\frac{d^3p_B}{(2\pi)^32p_B^0} \vert{\cal M}_n  \vert^{2} \nonumber \\
    & \times (2\pi)^4\delta(n\omega-p_A^0-p_B^0)\delta^3(\vec p_A + \vec p_B).
\end{align}
Notice that $n\leq 0$ does not contribute due to the energy conservation.
From this, for instance, the $\chi$ number density production rate through $\phi\phi\to \chi\chi$ can be computed by
\begin{align}
    R^{\rm (N)}_{\phi\phi\to \chi \chi} &= \sum_{n=1}^\infty \int dP^{(n)}_{\phi\phi\to \chi\chi},
\end{align}
which is the right hand side of the Boltzmann equation given by
\begin{align}
    \dot n_\chi + 3Hn_\chi &= R^{\rm (N)}_{\phi\phi\to \chi \chi}
\end{align}
with $\dot n_\chi\equiv dn_\chi/dt$.
Note that on the right-hand side of the above equation, we should multiply by 2 as two $\chi$ particles are produced per reaction and should multiply by 1/2 of the symmetry factor as the final state $\chi$'s are identical particles, resulting in cancellation of these factors.

Since $V\propto \phi^2$, $\phi$ is identified as a simple harmonic oscillator whose solution may be written as $\phi=\phi_0\cos(m_\phi t)$ with an amplitude $\phi_0$.
Given that the energy density of $\phi$ is given by $\rho_\phi = V(\phi_0)$ and $V_n = \rho_\phi/4$ for $n=2$, we end up with
\begin{align}
    R^{\rm (N)}_{\phi\phi\to \chi \chi} &=
    \frac{\rho_\phi^2}{128\pi M_P^4}\left(1+\frac{m_\chi^2}{2m_\phi^2}\right)^2\sqrt{1-\frac{m_\chi^2}{m_\phi^2}},
\end{align}
which is consistent with the results in Refs. \cite{Mambrini:2021zpp,Clery:2021bwz}

To find the distribution function of the final state particle $A$ produced through $\phi\phi\to A(p_A) B(p_B)$, we should solve the Boltzmann equation given by 
\begin{align}
    \frac{\partial f_A}{\partial t} - H |\vec p_A|\frac{\partial f_A}{\partial |\vec p_A|} = C[f_A]
\end{align}
with the collision term
\begin{align}
    C[f_A] &=
    \sum_{n=1}^\infty \frac{1}{2p_A^0}\int\frac{d^3p_B}{(2\pi)^32p_B^0}|{\cal M}_n|^2\nonumber\\
    &\qquad \times (2\pi)^4\delta(n\omega-p_A^0-p_B^0)\delta(\vec p_A+\vec p_B),\nonumber \\
    &=
    \sum_{n=1}^{\infty}\frac{\pi}{(n\omega)^2\beta_n}|{\cal M}_n|^2\delta\left(|\vec p_A|-\frac{n\omega}{2}\beta_n\right),
\end{align}
where
\begin{align}
    \beta_n \equiv \sqrt{\left(1-\frac{(m_A+m_B)^2}{(n\omega)^2}\right)\left(1-\frac{(m_A-m_B)^2}{(n\omega)^2}\right)}.
\end{align}
When the backward reaction $AB\to \phi\phi$ is negligible and $f_A,f_B\ll 1$, the Boltzmann equation has a generic solution given by 
\begin{align}
    f_A(|\vec p_A|,t) &=
    \int_{t_r}^tdt' C[f_A](|\vec p_A|\times a(t)/a(t'),t'),
    \label{eq:f_general}
\end{align}
where $a$ is the scale factor and $t_r$ is an arbitrary reference time \cite{Garcia:2018wtq,Ballesteros:2020adh,Garcia:2021iag}.
By plugging Eq.~\eqref{eq:M_n} into $C[f_\chi]$, $f_\chi$ is readily obtained as
\begin{align}
    f_\chi(|\vec p_\chi|,t) =&
    \frac{9\pi}{64}\left(\frac{H_e}{m_\phi}\right)^3\left(\frac{m_\phi}{|\vec p_\chi|(a/a_e)}\right)^{9/2} \nonumber 
    \\ & \times \left(1-\frac{m_\chi^2}{m_\phi^2}\right)^{5/4}\left(1+\frac{m_\chi^2}{2m_\phi^2}\right)^2,
    \label{eq:f_Boltzmann}
\end{align}
where we have used $3M_P^2H^2\simeq \rho_\phi = \rho_e(a/a_e)^{-3}$ with $a_e$ the scale factor at which the inflation ends and $\rho_e$ is the energy density at $a=a_e$.
In the same manner, $H_e$ is the Hubble parameter at $a=a_e$.
Notice that $|\vec p_\chi|(a/a_e)$ is a comoving momentum and a constant of time.

\section{The higher mode spectrum in the Bogoliubov approach}
\label{sec:Bogoliubov_high-k}

Instead of utilizing the transition probability to describe the particle production, in the Bogoliubov approach, one may deal with the time evolution of the wave function of the produced particle while keeping the effect of curved spacetime.
To do so, it is convenient to introduce the conformal time $\eta$ as $ad\eta=dt$ so that the metric is given as $g_{\mu\nu}=a^2\eta_{\mu\nu}$.

Integrating by parts, we may rewrite the action of $\chi$ as
\begin{align}
    S_\chi &=
    \int d^4x\left[\frac{1}{2}(\widetilde\chi')^2-\frac{1}{2}\widetilde\chi \omega^2\widetilde\chi\right],\\
    \omega^2 &\equiv
    -\nabla^2 + a^2m_\chi^2 + \Delta,
    \label{eq:def_omega}\\
    \Delta &\equiv -\frac{a''}{a} = \frac{1}{6}a^2 R,
    \label{eq:def_Delta}
\end{align}
where $\widetilde\chi$ is a rescaled field given by  $\chi=a^{-1}\widetilde\chi$, and $'\equiv d/d\eta$.
The equation of motion is thus given by
\begin{align}
    \widetilde\chi'' + \omega^2\widetilde\chi = 0.
    \label{eq:EOM_chi}
\end{align}
In this field definition, the conjugate momentum of $\widetilde\chi$ is defined by $\widetilde\pi\equiv \partial{\cal L}_\chi/\partial\widetilde\chi'=\widetilde\chi'$ so that the Hamiltonian is given by
\begin{align}
   H &=
   \int d^3x 
   \left[ \widetilde\pi \widetilde \chi' - {\cal L}  \right]
   =
   \int d^3x \left[\frac{1}{2}\widetilde\pi^2 + \frac{1}{2}\widetilde{\chi}\omega^2\widetilde{\chi}\right].
\end{align}
From Eq.~\eqref{eq:def_omega}, it is clear that the Hamiltonian is changing with time through the time dependence in $\omega$.
Therefore, we cannot decompose $\widetilde\chi$ based on the positive/negative frequency in the Fourier space as is normally done in the flat spacetime quantum field theory.
Instead, for the mode function $\widetilde\chi_k$ defined by
\begin{align}
    \widetilde\chi(x) &=
    \int \frac{d^3k}{(2\pi)^{3/2}}e^{i\vec k\cdot\vec x}\widetilde\chi_k,\\
    \widetilde\chi_k &=
    a_{\vec k}u_k + a^\dagger_{-\vec k}u_k^*
    \label{eq:u_k}
\end{align}
with $a_{\vec k}$ and $a_{\vec k}^\dagger$ being annihilation and creation operators, respectively, we decompose $u_k$ as
\begin{align}
    u_k &=
    \frac{A_k}{\sqrt{2\omega_k}}e^{-i\int\omega_k d\eta} + \frac{B_k}{\sqrt{2\omega_k}}e^{i\int\omega_k d\eta},
\end{align}
where $\omega_k^2=k^2+a^2m_\chi^2+\Delta$.
Time-dependent coefficients introduced in this way (here $A_k$ and $B_k$) are called Bogoliubov coefficients.
For latter convenience, we define
\begin{align}
    &
    \alpha_k \equiv A_k e^{-i\int\omega_k d\eta},\;\;\;
    \beta_k \equiv B_k e^{i\int\omega_k d\eta},
\end{align}
with which we parametrize the conjugate momentum as
\begin{align}
    \widetilde \pi(x) &= \int \frac{d^3k}{(2\pi)^{3/2}}e^{i\vec k\cdot\vec x}\widetilde\pi_k,
    \label{eq:conjugate_momentum1}
    \\
    \widetilde\pi_k&= a_{\vec k}v_k + a^\dagger_{-\vec k}v_k^*,\;\;\;
    v_k = \frac{-i\omega_k \alpha_k}{\sqrt{2\omega_k}} + \frac{i\omega_k\beta_k}{\sqrt{2\omega_k}}.
    \label{eq:conjugate_momentum2}
\end{align}
To satisfy the equation of motion $\widetilde\pi_k'+\omega_k^2\widetilde\chi_k=0$ as well as the definition $\widetilde\pi_k=\widetilde\chi_k'$, the coefficients are required to obey
\begin{align}
    \alpha'_k &= 
    -i\omega_k\alpha_k + \frac{\omega'_k}{2\omega_k}\beta_k,\label{eq:ode_alpha}\\
    \beta'_k &= 
    i\omega_k\beta_k + \frac{\omega'_k}{2\omega_k}\alpha_k.\label{eq:ode_beta}
\end{align}
Note that we have not assumed any slowly varying nature for $A_k$ and $B_k$ nor for $\alpha_k$ and $\beta_k$. Their time-dependence is fully taken into account and no approximation is made.

Since the occupation number is given by $|\beta_k|^2$ (equivalent to $f_\chi(|\vec p_\chi|,t)$ in the Boltzmann approach), the particle production can be seen by solving these coupled differential equations for $\beta_k$.
By setting $\alpha_k=1$ and $\beta_k=0$ as the initial condition, $\alpha_k(\eta) = e^{-i\omega_k \eta}$ and $\beta_k(\eta)=0$ when $\omega_k'\simeq 0$, which corresponds to the case of ordinary plane wave solution in the flat spacetime, namely, non-existence of particles in the initial state.
In particular, if $|\beta_k|^2\ll 1$ all the time (which is the case of the gravitational production with large $k$), then we may write 
\begin{align}
    \beta_k &\simeq
    \int_{\eta_i}^\eta d\eta'\frac{\omega'_k}{2\omega_k}e^{-2i\Omega_k(\eta')},\label{eq:beta_approximate}\\
    \Omega_k(\eta) &\equiv
    \int_{\eta_i}^\eta d\eta'\omega_k(\eta'),
\end{align}
where $t_i$ is taken as some reference time where the initial condition is set.

As the gravitational effect is induced by the background field dynamics through $\Delta$, we need to take a closer look at the inflaton dynamics during reheating.
By solving the equation of motion with the Hubble friction term,
\begin{align}
    \ddot\phi+3H\dot\phi+m_\phi^2\phi=0,
\end{align}
we find a solution of the form
\begin{align}
    \phi(t) = \frac{\phi_e}{m_\phi t}\sin(m_\phi t)
    \label{eq:phi_sol}
\end{align}
where $\phi_e$ is the inflaton amplitude at the end of inflation, and we have assumed $H\simeq 3/2t$.\footnote{This corresponds to taking the equation of state parameter $w\simeq 0$. However, since the inflaton pressure is quickly oscillating with time as will be seen later, this assumption is not entirely a good approximation. Nevertheless, we have numerically checked that the solution given in Eq. (\ref{eq:phi_sol}) is in good agreement with numerical computation.}
By identifying $\phi_0\equiv \phi_e/m_\phi t$, the slowly varying component of energy density, $\bar\rho_\phi$, is given by $\bar\rho_\phi=V(\phi_0) $, and accordingly, we define $ 3M_P^2\bar H^2 \equiv\bar\rho_\phi $. It should be noted that the fast oscillating mode in $\rho_\phi$ quickly becomes subdominant as time increases, whereas in the pressure, it is always the dominant contribution as the slowly varying component gives zero pressure.
This fast oscillating component in the pressure plays a central role in gravitational production.

At leading order, we obtain
\begin{align}
    \Delta &= \frac{1}{6}a^2R \simeq -\frac{1}{2}(a\bar H)^2[1-3\cos(2m_\phi t)],\\
    \Delta' & \simeq -(a\bar H)^3\left[\frac{\Delta}{(a\bar H)^2} - 3\frac{m_\phi}{\bar H}\sin(2m_\phi t)\right],
\end{align}
which are used in evaluating Eq. (\ref{eq:beta_approximate}) to find
\begin{widetext}
\begin{align}
    \beta_{k\gg a\bar H} &\simeq
    \frac{1}{2}
    \int_{\eta_i}^\eta d\eta'\left[\frac{a^3 \bar H m_\chi^2}{k^2+a^2m_\chi^2} + \frac{3}{2}\frac{(a\bar H)^3(m_\phi/\bar H)}{k^2+a^2m_\chi^2}\left(1+\frac{m_\chi^2}{2m_\phi^2}\right)\sin(2m_\phi t)\right]e^{-2i\Omega_k(\eta')}
    \label{eq:beta_approximate_khigh}
\end{align}
\end{widetext}
for $k\gg a\bar H$.
The integrand of the first term in Eq. (\ref{eq:beta_approximate_khigh}) is highly oscillating with $e^{-2i\Omega_k(\eta')}\simeq e^{-2i(k^2+a^2m_\chi^2)^{1/2}\eta'}$ leading to a delta function-like behavior ($\delta(k)$) for sufficiently large $\eta'$.
However, as we are considering modes with $k\gg a\bar H$, there is no solution for $k$ to satisfy the delta function, so we may drop this contribution.
For the second term in Eq. (\ref{eq:beta_approximate_khigh}), using $\sin(2m_\phi t)=(-i/2)(e^{2im_\phi t}-e^{-2im_\phi t})$, we obtain
\begin{align}
    \beta_{k\gg a\bar H} &\simeq
    i\frac{3}{8} \int_{\eta_i}^\eta d\eta' (a\bar H)^3\frac{m_\phi/\bar H}{k^2+a^2m_\chi^2}\left(1+\frac{m_\chi^2}{2m_\phi^2}\right)\nonumber\\
    &\times\left(e^{i\varphi_+} - e^{-i\varphi_-}\right),\\
    \varphi_\pm(\eta) &\equiv
    \pm 2m_\phi t - 2\Omega_k(\eta).
\end{align}
To perform the integration, we use the stationary phase approximation by expanding $\varphi_\pm$ as 
\begin{align}
    \varphi_\pm(\eta) &\simeq
    \varphi_\pm(\eta_k) + \varphi'_\pm(\eta_k)(\eta-\eta_k) + \frac{1}{2}\varphi''_\pm(\eta_k)(\eta-\eta_k)^2\nonumber\\
    &+(\text{higher order in} \;\;\eta-\eta_k),
\end{align}
where $\eta_k$ is defined to satisfy $\varphi'_\pm(\eta_k)=0$.
It is clear that $\varphi_-$ does not have a solution to satisfy this condition, so we drop the corresponding contribution in $\beta_{k\gg a\bar H}$, and $\varphi'_+(\eta_k)=0$ results in 
\begin{align}
    a_km_\phi = \omega_k(\eta_k) \simeq \sqrt{k^2 + a_k^2 m_\chi^2}
    &
    \;\;\;\Rightarrow\;\;\;
    a_k \simeq \sqrt{\frac{k^2}{m_\phi^2-m_\chi^2}}
    \label{eq:a_k}
\end{align}
with $a_k\equiv a(\eta_k)$.
Note that we require that $\eta_i < \eta_k < \eta$, otherwise the integrand is highly oscillating and is strongly damped to yield null contribution.
We also obtain
\begin{align}
    \varphi''_+(\eta_k) &\simeq 
    2k^2\frac{\bar H_k}{m_\phi},
\end{align}
where $\bar H_k \equiv \bar H(a_k)$.

Notice that $e^{i\varphi_+}$ highly oscillates except for the vicinity at $\eta=\eta_k$ when $\varphi''(\eta_k)\gg 1/(\eta-\eta_k)^2$, and thus the integrand with $\eta$ far from $\eta_k$ cancels in the integration, leaving the integrand with $\eta=\eta_k$.
Therefore, we obtain
\begin{align}
    \beta_{k\gg a\bar H} 
    &\simeq
    i\frac{3}{8}(a_k\bar H_k)^3\frac{m_\phi/\bar H_k}{k^2+a_k^2m_\chi^2}\left(1+\frac{m_\chi^2}{2m_\phi^2}\right)\nonumber\\
    &\times \sqrt{\frac{2\pi}{|\varphi_+''(\eta_k)|}}e^{\pm i\pi/4 + i\varphi_+(\eta_k)}.
\end{align}
With the normalization $a_e=1$ and thus $\bar H_k=H_ea_k^{-3/2}$, we end up with 
\begin{align}
    |\beta_{k\gg a\bar H}|^2 \simeq & 
    \frac{9\pi}{64}\left(\frac{H_e}{m_\phi}\right)^3\left(\frac{m_\phi}{k}\right)^{9/2}
    \\
    & \times \left(1-\frac{m_\chi^2}{m_\phi^2}\right)^{5/4}\left(1+\frac{m_\chi^2}{2m_\phi^2}\right)^2,
    \label{eq:beta_khigh_analytic}
\end{align}
which is identical to Eq. (\ref{eq:f_Boltzmann}) by noticing the comoving momentum $k=|\vec p_\chi|(a/a_e)$.
Therefore, we have confirmed that for $k\gg a_e H_e$, both the Boltzman and Bogoliubov approaches give the identical spectrum for the gravitational production of $\chi$.

\section{Vacuum contribution}
\label{sec:Bogoliubov_low-k}

In gravitational production, an important effect that cannot be caught by the Boltzmann approach is particle production due to the breaking down of the adiabaticity coming from the sudden change of the vacuum from the de-Sitter to the matter-dominated phase.
It turns out that $\chi$ with $k\ll H_e$ may be produced by this phase transition, and this production rate differs from what is obtained in the Boltzmann approach.
To compute $|\beta_{k\ll H_e}|^2$ for such modes, we need to develop the formalism further.
In doing so, we follow closely Ref. \cite{deGarciaMaia:1993ck};
up to Eq.~\eqref{end of review} is a review.

We set a reference time $\eta_r$ after which the Hubble parameter evolves as
\begin{align}
    H(a\geq a_r) &= H_r \left(\frac{a}{a_r}\right)^{-\frac{3(1+w_r)}{2}},
\end{align}
where $a_r\equiv a(\eta=\eta_r)$, $H_r\equiv H(\eta=\eta_r)$, and the equation of state parameter $w_r\equiv w(\eta=\eta_r)$.
Thus, the scale factor at $\eta\geq\eta_r$ may be written as
\begin{align}
    &
    a(\eta\geq \eta_r) =c_r(\eta - \bar\eta_r)^{q_r},\\
    &
    q_r \equiv \frac{2}{1+3w_r},\;\;\;
    c_r \equiv a_r\left(\frac{a_r H_r}{q_r}\right)^{q_r},\;\;\;
    \bar\eta_r \equiv \eta_r - \frac{q_r}{a_r H_r}.\label{eq:qr,cr,etarbar}
\end{align}
We also require that $a(\eta)$ and its first derivative $a'(\eta)$ are smooth at $\eta=\eta_r$.

Supposing that for the mode function $u_k(\eta)$ in Eq. (\ref{eq:u_k}) is given as $u_{r-1}$ for the time interval $\eta_{r-1}<\eta<\eta_r$ and as $u_r$ for $\eta_r < \eta < \eta_{r+1}$, the Bogoliubov coefficients can be computed by
\begin{align}
    &
    \alpha_k = -i W[u_{r-1}^*,u_r],\;\;\;\beta_k = i W[u_{r-1},u_r]
    \label{eq:Bogoliubov_Wronskian}
\end{align}
with the Wronskian $W[f,g]\equiv f'g - f g'$.
Note that $u_r$ is a solution of the equation of motion given by
\begin{align}
    &
    u_r'' + \omega_r^2u_r = 0,\label{eq:eom}\\
    &
    \omega_r^2 \equiv k^2 + c_r^2m_\chi^2(\eta-\bar\eta_r)^{2q_r} - \frac{q_r(q_r-1)}{(\eta-\bar\eta_r)^2}.
    \label{eq:omega_r}
\end{align}

The simplest application of this method is for the case of $m_\chi=0$.
We take $\eta_r\equiv \eta_e$ at which the transition from $w_{r-1}=-1$ (inflation) to $w_r=0$ (matter-domination) takes place.
Imposing the Bunch-Davies vacuum \cite{Chernikov:1968zm,Bunch:1978yq} as the initial condition,
\begin{align}
    u_{r-1}(\eta\to -\infty) &= \frac{e^{-ik\eta}}{\sqrt{2k}},
    \label{eq:BD-vacuum}
\end{align}
we find the solution for $u_{r-1}$ during inflation given by
\begin{align}
    u_{r-1} &= e^{i\theta_{r-1}}\sqrt{\frac{\pi}{4k}}\sqrt{y}H^{(2)}_{l_r}(y),
    \label{eq:u_r-1}
\end{align}
where $H^{(2)}_{l_r}(y)$ is the Hankel function of the second kind, $y\equiv k(\eta-\bar\eta_{r-1}), l_r\equiv q_{r-1}-1/2$, and $\theta_{r-1}$ denotes an irrelevant phase in the particle production and is dropped in the following expressions.
For $\eta>\eta_r$, $u_r$ has the same form as $u_{r-1}$ but replacing $y\to x\equiv k(\eta-\bar\eta_r), l_r\to m_r\equiv q_r - 1/2$, and $\theta_{r-1}\to \theta_r$.
Evaluating Eq.(\ref{eq:Bogoliubov_Wronskian}) at $\eta=\eta_r$, we obtain
\begin{align}
    \beta_k &= -i\frac{\pi}{4}\sqrt{\frac{q_{r-1}}{q_r}}x_r Q_\beta,\label{eq:Bogoliubov_massless_exact}\\
    Q_\beta &\equiv H^{(2)}_{l_r}(y_r)H^{(2)}_{m_r+1}(x_r) - H^{(2)}_{m_r}(x_r)H^{(2)}_{l_r+1}(y_r),
    \label{end of review}
\end{align}
where $x_r\equiv k(\eta_r-\bar\eta_r)=kq_r/a_rH_r, y_r\equiv k(\eta_r-\bar\eta_{r-1})=(q_{r-1}/q_r)x_r$.
By setting $a_r=a_e=1, H_r=H_e$ and taking $k\ll H_e$, we obtain
\begin{align}
    |\beta_{k\ll H_e}|^2 & \simeq \frac{9}{64}\left(\frac{H_e}{k}\right)^6.
    \label{eq:Bogoliubov_massless}
\end{align}

Finally, we consider the case when $m_\chi$ is not negligible compared to $H_e$.
In this case, for $u_{r-1}$, the solution remains the same form with\footnote{Note that Eq.~(\ref{eq:lr_massive}) should be taken as the definition of $q_{r-1}$, instead of that given in Eq.~(\ref{eq:qr,cr,etarbar}), namely, $q_{r-1}=-1/2-\sqrt{9/4-m_\chi^2/H_e^2}$, so that $u_{r-1}$ given in Eq.~(\ref{eq:u_r-1}) becomes the solution for Eq.~(\ref{eq:eom}).} 
\begin{align}
    l_r & \equiv -\sqrt{\frac{9}{4}-\frac{m_\chi^2}{H_r^2}}.\label{eq:lr_massive}
\end{align}
For $\eta>\eta_r$, we may approximate $\omega_r^2\simeq k^2 + c_r^2m_\chi^2(\eta-\bar\eta_r)^{2q_r}$ at later times as the third term in Eq. (\ref{eq:omega_r}) is supposed to become negligible soon, compared to the second term for $q_r=2$ ($w_r=0$).
By noticing that $\omega_r$ is monotonically increasing with $\eta$, we may approximate the solution by the WKB-like form given as
\begin{align}
    &
    u_r \simeq \frac{e^{-i\Omega_r}}{\sqrt{2\omega_r}},\;\;\;
    \Omega_r(\eta) \equiv \int_{\eta_r}^\eta d\eta'\omega_r(\eta).
\end{align}
Again, using Eq. (\ref{eq:Bogoliubov_Wronskian}), we obtain
\begin{widetext}
\begin{align}
    \beta_k & \simeq
    \frac{i}{2}\sqrt{\frac{\pi}{2k}}\left[ \left\{\frac{(1+2l_r)k}{2\sqrt{y_r\omega_r}} + \frac{\sqrt{y_r}\omega_r'}{2\omega_r^{3/2}}+i\sqrt{y_r\omega_r}\right\}H^{(2)}_{l_r}(y_r) - k\sqrt{\frac{y_r}{\omega_r}}H^{(2)}_{l_r+1}(y_r)\right],
    \label{eq:Bogoliubov_massive_exact}
\end{align}
\end{widetext}
where $\omega_r$ is evaluated at $x=x_r$.
This is one of our main results.
Note that irrelevant $e^{-i\Omega_r}$ factor is dropped in Eq. (\ref{eq:Bogoliubov_massive_exact}) as was done in the massless case.
Thus, for $k\ll H_e$, the asymptotic form of $|\beta_k|^2$ becomes
\begin{align}
    |\beta_{k\ll H_e}|^2 \simeq \frac{9}{32}\left(\frac{H_e}{m_\chi}\right)\left(\frac{H_e}{k}\right)^3.
    \label{eq:Bogoliubov_massive}
\end{align}

\begin{figure*}[t!]
    \centering
    \includegraphics[width=0.95\textwidth]{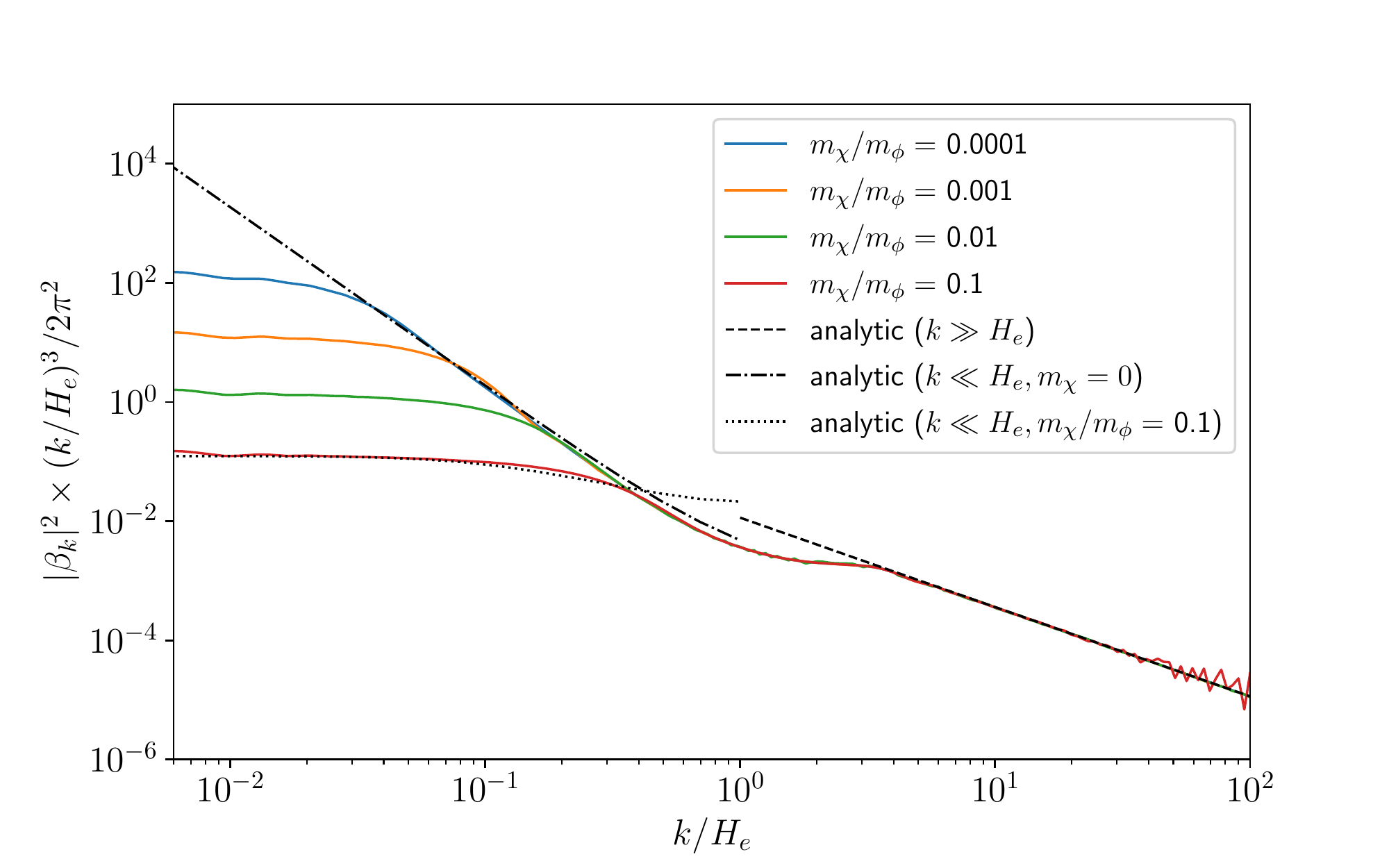}
    \caption{Spectrum of gravitationally produced $\chi$ obtained by computing both numerically and analytically. The solid lines are obtained by numerical computation. The spectrum for $k\gg H_e$ is $|\beta_k|^2\propto 1/k^{9/2}$ and well fitted by the analytic result in Eq. (\ref{eq:f_Boltzmann}) [or equivalently Eq. (\ref{eq:beta_khigh_analytic})] shown in the dashed line. The dot-dashed line shows Eq. (\ref{eq:Bogoliubov_massless_exact}) which is valid only for $k\ll H_e$ and $m_\chi=0$. The dotted line is obtained by Eq. (\ref{eq:Bogoliubov_massive_exact}) which is also valid only for $k\ll H_e$.}
    \label{fig:summary}
\end{figure*}

\section{Summary and discussion}
\label{sec:summary}

Gravitational particle production is a minimal contribution to reheating the Universe after the end of inflation. 
To study this production channel, two different approaches have commonly been considered, one of which is based on the Boltzmann equation, and the other is based on the Bogoliubov transformation.
However, their (in)equivalence in particle production is not obvious.
We have explicitly shown by looking at the particle spectrum that for the higher $k$-modes, both approaches are equivalent.
For the lower $k$-modes, however, the Boltzmann approach can not deal with the particle production.
By developing a theoretical framework, we have obtained the analytic estimate of such contribution based on the Bogoliubov approach.

Our main results are summarized in Fig. \ref{fig:summary} where the solid lines are obtained by numerically solving the equation of motion Eq. (\ref{eq:EOM_chi}) and computing
\begin{align}
    &
    |\beta_k|^2 = \frac{|\widetilde\chi_k'|^2}{2\omega_k} + \frac{\omega_k}{2}|\widetilde\chi_k|^2 + \frac{i}{2}W[\widetilde\chi^*,\widetilde\chi]
\end{align}
in the limit of $\eta\to\infty$ at which the energy density is well defined ($\omega_k^2>0)$.
The initial condition for $\widetilde\chi_k$ is taken as the Bunch-Davies vacuum given in Eq. (\ref{eq:BD-vacuum}).
For the numerical results, we take $H_e=m_\phi$, which just sets the scale of the horizontal axis in the figure.

The spectrum at higher $k-$modes ($k\gg H_e$) is well explained by both Boltzmann and Bogoliubov approaches given in Eqs. (\ref{eq:f_Boltzmann}) and (\ref{eq:beta_khigh_analytic}), respectively.
The analytic estimate is shown in the dashed line in the figure, while we take $H_e=m_\phi/1.25$, instead of $H_e=m_\phi$, to fit the numerical result.
As one might notice, there is no proper way to define $H_e$, as the transition from de Sitter to the matter-domination phase is smooth, rather than discrete.
Therefore, the analytic estimate does not exactly coincide with the numerical one in general.

For the lower $k-$modes ($k\ll H_e$), the analytic estimate for $m_\chi=0$ given by Eq. (\ref{eq:Bogoliubov_massless_exact}) and $m_\chi=m_\phi/10$ given by Eq. (\ref{eq:Bogoliubov_massive_exact}) are shown in dot-dashed and dotted lines, respectively.
For the massless case, we again take $H_e=m_\phi/1.25$, while $H_e=m_\phi$ for the massive case.
Both are in good agreement with the numerical results.

Notice that for $m_\chi\ll m_\phi$, there is an infrared divergence since, for instance, the energy density scales as $|\beta_k|^2 k^4\propto 1/k^2$ for $k\ll H_e$ in the massless case.
Therefore, we need a cutoff in the deep infrared regime, for which one may take the wave number, $k_{\rm CMB}$, taken in the cosmic microwave background measurements.
By writing $a/a_e\equiv e^{-N}$ during inflation, $k_{\rm CMB}$ may roughly be set by $k_{\rm CMB}/H_e=\sqrt{2-m_\chi^2/H_e^2}e^{-N}$ where $N$ depends on the inflationary model, for which our choice is $N=55$ throughout our analysis.

It should be noted that both Eqs. (\ref{eq:Bogoliubov_massless_exact}) and (\ref{eq:Bogoliubov_massive_exact}) can not be applied for $k\gg H_e$ since such modes always stay well inside the horizon, i.e, remain the plane wave as if the particles were in the Minkowski spacetime, and thus do not contribute particle production.
In addition, for the massive case, we have shown only $m_\chi=m_\phi/10$.
This is because the approximation used in deriving Eq. (\ref{eq:Bogoliubov_massive_exact}) becomes worse when $0< m_\chi \ll m_\phi$.
In such cases, in $\omega_r^2$ given in Eq. (\ref{eq:omega_r}), there exists a time scale where the third term is still dominant, though the second term (mass term) will dominate in the distant future.
Therefore, to observe the finite mass effect, for instance, one may need to divide the time scales into three, namely, during inflation, the time scale when the mass term in $\omega_r$ can still be negligible, and that when the mass term becomes dominant.

Our results can be used for various phenomenological studies.
For instance, one may consider the particle production out of the vacuum when the inflaton potential is not quadratic during reheating.
In such cases, the equation of state during reheating can be different than $w=0$, and thus the resulting spectrum can be modified.
Also, we have not specified the particle nature of $\chi$ for which one may identify a Higgs(-like) particle whose decay generates dark matter.
In such scenarios, the spectrum of the parent particle is necessary to compute the dark matter abundance, which can be investigated in our framework.
Furthermore, one may include the non-minimal coupling of $\chi$ to the Ricci scalar, extend the potential of $\chi$ beyond the quadratic, or even consider different spins of $\chi$, which we will leave for future works.

{\bf Note added:} After completion of the draft and while preparing for submission, a paper~\cite{Garcia:2022vwm} appeared that partially overlapped with our analysis of gravitational particle production, in obtaining a numerical result from the Bogoliubov approach and an analytical result from the Boltzmann approach. In this work, we have obtained all the relevant results and have compared Boltzmann and Bogoliubov approaches both numerically and analytically.

\section*{Acknowledgements}
We would like to thank S. Clery, M. A. G. Garcia, Y. Mambrini, and S. Verner for the useful discussion.
This work is in part supported by the JSPS KAKENHI Grant Nos.\ 19H01899 (KK and KO) and 21H01107 (KO). 
KK acknowledges support by Institut Pascal at Université Paris-Saclay during the Paris-Saclay Astroparticle Symposium 2021, with the support of the P2IO Laboratory of Excellence (program “Investissements d’avenir” ANR-11-IDEX-0003-01 Paris-Saclay and ANR-10-LABX-0038), the P2I axis of the Graduate School Physics of Université Paris-Saclay, as well as IJCLab, CEA, IPhT, APPEC, the IN2P3 master projet UCMN and EuCAPT.
The work of SML was supported by National Research Foundation (NRF) grants funded by the Korean government (NRF-2020R1A6A3A13076216). SML is also supported by the Hyundai Motor Chung Mong-Koo Foundation Scholarship.

\bibliographystyle{utphys}
\bibliography{biblio}

\providecommand{\href}[2]{#2}\begingroup\raggedright\begin{thebibliography}{100}

\bibitem{Olive:1989nu}
K.~A. Olive, ``{Inflation},''
  \href{http://dx.doi.org/10.1016/0370-1573(90)90144-Q}{{\em Phys. Rept.}
  {\bfseries 190} (1990) 307--403}.

\bibitem{Linde:1990flp}
A.~D. Linde, {\em {Particle physics and inflationary cosmology}}, vol.~5 of
  {\em Contemp. Concepts Phys.}
\newblock Harwood, 1990.
\newblock \href{http://arxiv.org/abs/hep-th/0503203}{{\ttfamily
  hep-th/0503203}}.

\bibitem{Lyth:1998xn}
D.~H. Lyth and A.~Riotto, ``{Particle physics models of inflation and the
  cosmological density perturbation},''
  \href{http://dx.doi.org/10.1016/S0370-1573(98)00128-8}{{\em Phys. Rept.}
  {\bfseries 314} (1999) 1--146},
  \href{http://arxiv.org/abs/hep-ph/9807278}{{\ttfamily arXiv:hep-ph/9807278}}.

\bibitem{Linde:2000kn}
A.~D. Linde, ``{Inflationary cosmology},''
  \href{http://dx.doi.org/10.1016/S0370-1573(00)00038-7}{{\em Phys. Rept.}
  {\bfseries 333} (2000) 575--591}.

\bibitem{Martin:2013tda}
J.~Martin, C.~Ringeval, and V.~Vennin, ``{Encyclop\ae{}dia Inflationaris},''
  \href{http://dx.doi.org/10.1016/j.dark.2014.01.003}{{\em Phys. Dark Univ.}
  {\bfseries 5-6} (2014) 75--235},
  \href{http://arxiv.org/abs/1303.3787}{{\ttfamily arXiv:1303.3787
  [astro-ph.CO]}}.

\bibitem{Martin:2013nzq}
J.~Martin, C.~Ringeval, R.~Trotta, and V.~Vennin, ``{The Best Inflationary
  Models After Planck},''
  \href{http://dx.doi.org/10.1088/1475-7516/2014/03/039}{{\em JCAP} {\bfseries
  03} (2014) 039}, \href{http://arxiv.org/abs/1312.3529}{{\ttfamily
  arXiv:1312.3529 [astro-ph.CO]}}.

\bibitem{Martin:2015dha}
J.~Martin, ``{The Observational Status of Cosmic Inflation after Planck},''
  \href{http://dx.doi.org/10.1007/978-3-319-44769-8_2}{{\em Astrophys. Space
  Sci. Proc.} {\bfseries 45} (2016) 41--134},
  \href{http://arxiv.org/abs/1502.05733}{{\ttfamily arXiv:1502.05733
  [astro-ph.CO]}}.

\bibitem{Alpher:1948ve}
R.~A. Alpher, H.~Bethe, and G.~Gamow, ``{The origin of chemical elements},''
  \href{http://dx.doi.org/10.1103/PhysRev.73.803}{{\em Phys. Rev.} {\bfseries
  73} (1948) 803--804}.

\bibitem{Walker:1991ap}
T.~P. Walker, G.~Steigman, D.~N. Schramm, K.~A. Olive, and H.-S. Kang,
  ``{Primordial nucleosynthesis redux},''
  \href{http://dx.doi.org/10.1086/170255}{{\em Astrophys. J.} {\bfseries 376}
  (1991) 51--69}.

\bibitem{Olive:1999ij}
K.~A. Olive, G.~Steigman, and T.~P. Walker, ``{Primordial nucleosynthesis:
  Theory and observations},''
  \href{http://dx.doi.org/10.1016/S0370-1573(00)00031-4}{{\em Phys. Rept.}
  {\bfseries 333} (2000) 389--407},
  \href{http://arxiv.org/abs/astro-ph/9905320}{{\ttfamily
  arXiv:astro-ph/9905320}}.

\bibitem{Fields:2006bzp}
B.~D. Fields and K.~A. Olive, ``{Big bang nucleosynthesis},''
  \href{http://dx.doi.org/10.1016/j.nuclphysa.2004.10.033}{{\em Nucl. Phys. A}
  {\bfseries 777} (2006) 208--225}.

\bibitem{Fields:2014uja}
B.~D. Fields, P.~Molaro, and S.~Sarkar, ``{Big-Bang Nucleosynthesis},'' {\em
  Chin. Phys. C} {\bfseries 38} (2014) 339--344,
  \href{http://arxiv.org/abs/1412.1408}{{\ttfamily arXiv:1412.1408
  [astro-ph.CO]}}.

\bibitem{Steigman:2007xt}
G.~Steigman, ``{Primordial Nucleosynthesis in the Precision Cosmology Era},''
  \href{http://dx.doi.org/10.1146/annurev.nucl.56.080805.140437}{{\em Ann. Rev.
  Nucl. Part. Sci.} {\bfseries 57} (2007) 463--491},
  \href{http://arxiv.org/abs/0712.1100}{{\ttfamily arXiv:0712.1100
  [astro-ph]}}.

\bibitem{Cyburt:2001pp}
R.~H. Cyburt, B.~D. Fields, and K.~A. Olive, ``{The NACRE thermonuclear
  reaction compilation and big bang nucleosynthesis},''
  \href{http://dx.doi.org/10.1016/S1384-1076(01)00053-7}{{\em New Astron.}
  {\bfseries 6} (2001) 215--238},
  \href{http://arxiv.org/abs/astro-ph/0102179}{{\ttfamily
  arXiv:astro-ph/0102179}}.

\bibitem{Nollett:2000fh}
K.~M. Nollett and S.~Burles, ``{Estimating reaction rates and uncertainties for
  primordial nucleosynthesis},''
  \href{http://dx.doi.org/10.1103/PhysRevD.61.123505}{{\em Phys. Rev. D}
  {\bfseries 61} (2000) 123505},
  \href{http://arxiv.org/abs/astro-ph/0001440}{{\ttfamily
  arXiv:astro-ph/0001440}}.

\bibitem{Burles:2000zk}
S.~Burles, K.~M. Nollett, and M.~S. Turner, ``{Big bang nucleosynthesis
  predictions for precision cosmology},''
  \href{http://dx.doi.org/10.1086/320251}{{\em Astrophys. J. Lett.} {\bfseries
  552} (2001) L1--L6}, \href{http://arxiv.org/abs/astro-ph/0010171}{{\ttfamily
  arXiv:astro-ph/0010171}}.

\bibitem{Vangioni-Flam:2000yfu}
E.~Vangioni-Flam, A.~Coc, and M.~Casse, ``{Big bang nucleosynthesis updated
  with the nacre compilation},'' {\em Astron. Astrophys.} {\bfseries 360}
  (2000) 15, \href{http://arxiv.org/abs/astro-ph/0002248}{{\ttfamily
  arXiv:astro-ph/0002248}}.

\bibitem{Cyburt:2001pq}
R.~H. Cyburt, B.~D. Fields, and K.~A. Olive, ``{Primordial nucleosynthesis with
  CMB inputs: Probing the early universe and light element astrophysics},''
  \href{http://dx.doi.org/10.1016/S0927-6505(01)00171-2}{{\em Astropart. Phys.}
  {\bfseries 17} (2002) 87--100},
  \href{http://arxiv.org/abs/astro-ph/0105397}{{\ttfamily
  arXiv:astro-ph/0105397}}.

\bibitem{Cyburt:2003fe}
R.~H. Cyburt, B.~D. Fields, and K.~A. Olive, ``{Primordial nucleosynthesis in
  light of WMAP},''
  \href{http://dx.doi.org/10.1016/j.physletb.2003.06.026}{{\em Phys. Lett. B}
  {\bfseries 567} (2003) 227--234},
  \href{http://arxiv.org/abs/astro-ph/0302431}{{\ttfamily
  arXiv:astro-ph/0302431}}.

\bibitem{Coc:2003ce}
A.~Coc, E.~Vangioni-Flam, P.~Descouvemont, A.~Adahchour, and C.~Angulo,
  ``{Updated Big Bang nucleosynthesis confronted to WMAP observations and to
  the abundance of light elements},''
  \href{http://dx.doi.org/10.1086/380121}{{\em Astrophys. J.} {\bfseries 600}
  (2004) 544--552}, \href{http://arxiv.org/abs/astro-ph/0309480}{{\ttfamily
  arXiv:astro-ph/0309480}}.

\bibitem{Cuoco:2003cu}
A.~Cuoco, F.~Iocco, G.~Mangano, G.~Miele, O.~Pisanti, and P.~D. Serpico,
  ``{Present status of primordial nucleosynthesis after WMAP: results from a
  new BBN code},'' \href{http://dx.doi.org/10.1142/S0217751X04019548}{{\em Int.
  J. Mod. Phys. A} {\bfseries 19} (2004) 4431--4454},
  \href{http://arxiv.org/abs/astro-ph/0307213}{{\ttfamily
  arXiv:astro-ph/0307213}}.

\bibitem{Serpico:2004gx}
P.~D. Serpico, S.~Esposito, F.~Iocco, G.~Mangano, G.~Miele, and O.~Pisanti,
  ``{Nuclear reaction network for primordial nucleosynthesis: A Detailed
  analysis of rates, uncertainties and light nuclei yields},''
  \href{http://dx.doi.org/10.1088/1475-7516/2004/12/010}{{\em JCAP} {\bfseries
  12} (2004) 010}, \href{http://arxiv.org/abs/astro-ph/0408076}{{\ttfamily
  arXiv:astro-ph/0408076}}.

\bibitem{Cyburt:2004cq}
R.~H. Cyburt, ``{Primordial nucleosynthesis for the new cosmology: Determining
  uncertainties and examining concordance},''
  \href{http://dx.doi.org/10.1103/PhysRevD.70.023505}{{\em Phys. Rev. D}
  {\bfseries 70} (2004) 023505},
  \href{http://arxiv.org/abs/astro-ph/0401091}{{\ttfamily
  arXiv:astro-ph/0401091}}.

\bibitem{Descouvemont:2004cw}
P.~Descouvemont, A.~Adahchour, C.~Angulo, A.~Coc, and E.~Vangioni-Flam,
  ``{Compilation and R-matrix analysis of Big Bang nuclear reaction rates},''
  \href{http://dx.doi.org/10.1016/j.adt.2004.08.001}{{\em Atom. Data Nucl. Data
  Tabl.} {\bfseries 88} (2004) 203--236},
  \href{http://arxiv.org/abs/astro-ph/0407101}{{\ttfamily
  arXiv:astro-ph/0407101}}.

\bibitem{Iocco:2008va}
F.~Iocco, G.~Mangano, G.~Miele, O.~Pisanti, and P.~D. Serpico, ``{Primordial
  Nucleosynthesis: from precision cosmology to fundamental physics},''
  \href{http://dx.doi.org/10.1016/j.physrep.2009.02.002}{{\em Phys. Rept.}
  {\bfseries 472} (2009) 1--76},
  \href{http://arxiv.org/abs/0809.0631}{{\ttfamily arXiv:0809.0631
  [astro-ph]}}.

\bibitem{Pisanti:2007hk}
O.~Pisanti, A.~Cirillo, S.~Esposito, F.~Iocco, G.~Mangano, G.~Miele, and P.~D.
  Serpico, ``{PArthENoPE: Public Algorithm Evaluating the Nucleosynthesis of
  Primordial Elements},''
  \href{http://dx.doi.org/10.1016/j.cpc.2008.02.015}{{\em Comput. Phys.
  Commun.} {\bfseries 178} (2008) 956--971},
  \href{http://arxiv.org/abs/0705.0290}{{\ttfamily arXiv:0705.0290
  [astro-ph]}}.

\bibitem{Coc:2011az}
A.~Coc, S.~Goriely, Y.~Xu, M.~Saimpert, and E.~Vangioni, ``{Standard Big-Bang
  Nucleosynthesis up to CNO with an improved extended nuclear network},''
  \href{http://dx.doi.org/10.1088/0004-637X/744/2/158}{{\em Astrophys. J.}
  {\bfseries 744} (2012) 158}, \href{http://arxiv.org/abs/1107.1117}{{\ttfamily
  arXiv:1107.1117 [astro-ph.CO]}}.

\bibitem{Cyburt:2008kw}
R.~H. Cyburt, B.~D. Fields, and K.~A. Olive, ``{An Update on the big bang
  nucleosynthesis prediction for Li-7: The problem worsens},''
  \href{http://dx.doi.org/10.1088/1475-7516/2008/11/012}{{\em JCAP} {\bfseries
  11} (2008) 012}, \href{http://arxiv.org/abs/0808.2818}{{\ttfamily
  arXiv:0808.2818 [astro-ph]}}.

\bibitem{Coc:2014oia}
A.~Coc, J.-P. Uzan, and E.~Vangioni, ``{Standard big bang nucleosynthesis and
  primordial CNO Abundances after Planck},''
  \href{http://dx.doi.org/10.1088/1475-7516/2014/10/050}{{\em JCAP} {\bfseries
  10} (2014) 050}, \href{http://arxiv.org/abs/1403.6694}{{\ttfamily
  arXiv:1403.6694 [astro-ph.CO]}}.

\bibitem{Coc:2015bhi}
A.~Coc, P.~Petitjean, J.-P. Uzan, E.~Vangioni, P.~Descouvemont, C.~Iliadis, and
  R.~Longland, ``{New reaction rates for improved primordial D/H calculation
  and the cosmic evolution of deuterium},''
  \href{http://dx.doi.org/10.1103/PhysRevD.92.123526}{{\em Phys. Rev. D}
  {\bfseries 92} no.~12, (2015) 123526},
  \href{http://arxiv.org/abs/1511.03843}{{\ttfamily arXiv:1511.03843
  [astro-ph.CO]}}.

\bibitem{Pitrou:2018cgg}
C.~Pitrou, A.~Coc, J.-P. Uzan, and E.~Vangioni, ``{Precision big bang
  nucleosynthesis with improved Helium-4 predictions},''
  \href{http://dx.doi.org/10.1016/j.physrep.2018.04.005}{{\em Phys. Rept.}
  {\bfseries 754} (2018) 1--66},
  \href{http://arxiv.org/abs/1801.08023}{{\ttfamily arXiv:1801.08023
  [astro-ph.CO]}}.

\bibitem{Cyburt:2015mya}
R.~H. Cyburt, B.~D. Fields, K.~A. Olive, and T.-H. Yeh, ``{Big Bang
  Nucleosynthesis: 2015},''
  \href{http://dx.doi.org/10.1103/RevModPhys.88.015004}{{\em Rev. Mod. Phys.}
  {\bfseries 88} (2016) 015004},
  \href{http://arxiv.org/abs/1505.01076}{{\ttfamily arXiv:1505.01076
  [astro-ph.CO]}}.

\bibitem{Fields:2019pfx}
B.~D. Fields, K.~A. Olive, T.-H. Yeh, and C.~Young, ``{Big-Bang Nucleosynthesis
  after Planck},'' \href{http://dx.doi.org/10.1088/1475-7516/2020/03/010}{{\em
  JCAP} {\bfseries 03} (2020) 010},
  \href{http://arxiv.org/abs/1912.01132}{{\ttfamily arXiv:1912.01132
  [astro-ph.CO]}}. [Erratum: JCAP 11, E02 (2020)].

\bibitem{Yeh:2020mgl}
T.-H. Yeh, K.~A. Olive, and B.~D. Fields, ``{The impact of new $d(p,\gamma)$3
  rates on Big Bang Nucleosynthesis},''
  \href{http://dx.doi.org/10.1088/1475-7516/2021/03/046}{{\em JCAP} {\bfseries
  03} (2021) 046}, \href{http://arxiv.org/abs/2011.13874}{{\ttfamily
  arXiv:2011.13874 [astro-ph.CO]}}.

\bibitem{Fukugita:1986hr}
M.~Fukugita and T.~Yanagida, ``{Baryogenesis Without Grand Unification},''
  \href{http://dx.doi.org/10.1016/0370-2693(86)91126-3}{{\em Phys. Lett. B}
  {\bfseries 174} (1986) 45--47}.

\bibitem{Barbieri:1999ma}
R.~Barbieri, P.~Creminelli, A.~Strumia, and N.~Tetradis, ``{Baryogenesis
  through leptogenesis},''
  \href{http://dx.doi.org/10.1016/S0550-3213(00)00011-0}{{\em Nucl. Phys. B}
  {\bfseries 575} (2000) 61--77},
  \href{http://arxiv.org/abs/hep-ph/9911315}{{\ttfamily arXiv:hep-ph/9911315}}.

\bibitem{Davidson:2002qv}
S.~Davidson and A.~Ibarra, ``{A Lower bound on the right-handed neutrino mass
  from leptogenesis},''
  \href{http://dx.doi.org/10.1016/S0370-2693(02)01735-5}{{\em Phys. Lett. B}
  {\bfseries 535} (2002) 25--32},
  \href{http://arxiv.org/abs/hep-ph/0202239}{{\ttfamily arXiv:hep-ph/0202239}}.

\bibitem{Pilaftsis:2003gt}
A.~Pilaftsis and T.~E.~J. Underwood, ``{Resonant leptogenesis},''
  \href{http://dx.doi.org/10.1016/j.nuclphysb.2004.05.029}{{\em Nucl. Phys. B}
  {\bfseries 692} (2004) 303--345},
  \href{http://arxiv.org/abs/hep-ph/0309342}{{\ttfamily arXiv:hep-ph/0309342}}.

\bibitem{Raidal:2004vt}
M.~Raidal, A.~Strumia, and K.~Turzynski, ``{Low-scale standard supersymmetric
  leptogenesis},'' \href{http://dx.doi.org/10.1016/j.physletb.2005.01.040}{{\em
  Phys. Lett. B} {\bfseries 609} (2005) 351--359},
  \href{http://arxiv.org/abs/hep-ph/0408015}{{\ttfamily arXiv:hep-ph/0408015}}.
  [Addendum: Phys.Lett.B 632, 752--753 (2006)].

\bibitem{Nardi:2006fx}
E.~Nardi, Y.~Nir, E.~Roulet, and J.~Racker, ``{The Importance of flavor in
  leptogenesis},'' \href{http://dx.doi.org/10.1088/1126-6708/2006/01/164}{{\em
  JHEP} {\bfseries 01} (2006) 164},
  \href{http://arxiv.org/abs/hep-ph/0601084}{{\ttfamily arXiv:hep-ph/0601084}}.

\bibitem{Abada:2006ea}
A.~Abada, S.~Davidson, A.~Ibarra, F.~X. Josse-Michaux, M.~Losada, and
  A.~Riotto, ``{Flavour Matters in Leptogenesis},''
  \href{http://dx.doi.org/10.1088/1126-6708/2006/09/010}{{\em JHEP} {\bfseries
  09} (2006) 010}, \href{http://arxiv.org/abs/hep-ph/0605281}{{\ttfamily
  arXiv:hep-ph/0605281}}.

\bibitem{Buchmuller:2012wn}
W.~Buchmuller, V.~Domcke, and K.~Schmitz, ``{Spontaneous B-L Breaking as the
  Origin of the Hot Early Universe},''
  \href{http://dx.doi.org/10.1016/j.nuclphysb.2012.05.001}{{\em Nucl. Phys. B}
  {\bfseries 862} (2012) 587--632},
  \href{http://arxiv.org/abs/1202.6679}{{\ttfamily arXiv:1202.6679 [hep-ph]}}.

\bibitem{Khlebnikov:1988sr}
S.~Y. Khlebnikov and M.~E. Shaposhnikov, ``{The Statistical Theory of Anomalous
  Fermion Number Nonconservation},''
  \href{http://dx.doi.org/10.1016/0550-3213(88)90133-2}{{\em Nucl. Phys. B}
  {\bfseries 308} (1988) 885--912}.

\bibitem{Harvey:1990qw}
J.~A. Harvey and M.~S. Turner, ``{Cosmological baryon and lepton number in the
  presence of electroweak fermion number violation},''
  \href{http://dx.doi.org/10.1103/PhysRevD.42.3344}{{\em Phys. Rev. D}
  {\bfseries 42} (1990) 3344--3349}.

\bibitem{Kofman:1997yn}
L.~Kofman, A.~D. Linde, and A.~A. Starobinsky, ``{Towards the theory of
  reheating after inflation},''
  \href{http://dx.doi.org/10.1103/PhysRevD.56.3258}{{\em Phys. Rev. D}
  {\bfseries 56} (1997) 3258--3295},
  \href{http://arxiv.org/abs/hep-ph/9704452}{{\ttfamily arXiv:hep-ph/9704452}}.

\bibitem{Bassett:1998wg}
B.~A. Bassett, D.~I. Kaiser, and R.~Maartens, ``{General relativistic
  preheating after inflation},''
  \href{http://dx.doi.org/10.1016/S0370-2693(99)00478-5}{{\em Phys. Lett. B}
  {\bfseries 455} (1999) 84--89},
  \href{http://arxiv.org/abs/hep-ph/9808404}{{\ttfamily arXiv:hep-ph/9808404}}.

\bibitem{Felder:1998vq}
G.~N. Felder, L.~Kofman, and A.~D. Linde, ``{Instant preheating},''
  \href{http://dx.doi.org/10.1103/PhysRevD.59.123523}{{\em Phys. Rev. D}
  {\bfseries 59} (1999) 123523},
  \href{http://arxiv.org/abs/hep-ph/9812289}{{\ttfamily arXiv:hep-ph/9812289}}.

\bibitem{Greene:1998nh}
P.~B. Greene and L.~Kofman, ``{Preheating of fermions},''
  \href{http://dx.doi.org/10.1016/S0370-2693(99)00020-9}{{\em Phys. Lett. B}
  {\bfseries 448} (1999) 6--12},
  \href{http://arxiv.org/abs/hep-ph/9807339}{{\ttfamily arXiv:hep-ph/9807339}}.

\bibitem{Bassett:1999mt}
B.~A. Bassett, F.~Tamburini, D.~I. Kaiser, and R.~Maartens, ``{Metric
  preheating and limitations of linearized gravity. 2.},''
  \href{http://dx.doi.org/10.1016/S0550-3213(99)00495-2}{{\em Nucl. Phys. B}
  {\bfseries 561} (1999) 188--240},
  \href{http://arxiv.org/abs/hep-ph/9901319}{{\ttfamily arXiv:hep-ph/9901319}}.

\bibitem{Felder:1999pv}
G.~N. Felder, L.~Kofman, and A.~D. Linde, ``{Inflation and preheating in NO
  models},'' \href{http://dx.doi.org/10.1103/PhysRevD.60.103505}{{\em Phys.
  Rev. D} {\bfseries 60} (1999) 103505},
  \href{http://arxiv.org/abs/hep-ph/9903350}{{\ttfamily arXiv:hep-ph/9903350}}.

\bibitem{Chung:1999ve}
D.~J.~H. Chung, E.~W. Kolb, A.~Riotto, and I.~I. Tkachev, ``{Probing Planckian
  physics: Resonant production of particles during inflation and features in
  the primordial power spectrum},''
  \href{http://dx.doi.org/10.1103/PhysRevD.62.043508}{{\em Phys. Rev. D}
  {\bfseries 62} (2000) 043508},
  \href{http://arxiv.org/abs/hep-ph/9910437}{{\ttfamily arXiv:hep-ph/9910437}}.

\bibitem{Greene:2000ew}
P.~B. Greene and L.~Kofman, ``{On the theory of fermionic preheating},''
  \href{http://dx.doi.org/10.1103/PhysRevD.62.123516}{{\em Phys. Rev. D}
  {\bfseries 62} (2000) 123516},
  \href{http://arxiv.org/abs/hep-ph/0003018}{{\ttfamily arXiv:hep-ph/0003018}}.

\bibitem{Peloso:2000hy}
M.~Peloso and L.~Sorbo, ``{Preheating of massive fermions after inflation:
  Analytical results},''
  \href{http://dx.doi.org/10.1088/1126-6708/2000/05/016}{{\em JHEP} {\bfseries
  05} (2000) 016}, \href{http://arxiv.org/abs/hep-ph/0003045}{{\ttfamily
  arXiv:hep-ph/0003045}}.

\bibitem{Dufaux:2006ee}
J.~F. Dufaux, G.~N. Felder, L.~Kofman, M.~Peloso, and D.~Podolsky,
  ``{Preheating with trilinear interactions: Tachyonic resonance},''
  \href{http://dx.doi.org/10.1088/1475-7516/2006/07/006}{{\em JCAP} {\bfseries
  07} (2006) 006}, \href{http://arxiv.org/abs/hep-ph/0602144}{{\ttfamily
  arXiv:hep-ph/0602144}}.

\bibitem{Frolov:2010sz}
A.~V. Frolov, ``{Non-linear Dynamics and Primordial Curvature Perturbations
  from Preheating},''
  \href{http://dx.doi.org/10.1088/0264-9381/27/12/124006}{{\em Class. Quant.
  Grav.} {\bfseries 27} (2010) 124006},
  \href{http://arxiv.org/abs/1004.3559}{{\ttfamily arXiv:1004.3559 [gr-qc]}}.

\bibitem{Jedamzik:2010dq}
K.~Jedamzik, M.~Lemoine, and J.~Martin, ``{Collapse of Small-Scale Density
  Perturbations during Preheating in Single Field Inflation},''
  \href{http://dx.doi.org/10.1088/1475-7516/2010/09/034}{{\em JCAP} {\bfseries
  09} (2010) 034}, \href{http://arxiv.org/abs/1002.3039}{{\ttfamily
  arXiv:1002.3039 [astro-ph.CO]}}.

\bibitem{Amin:2014eta}
M.~A. Amin, M.~P. Hertzberg, D.~I. Kaiser, and J.~Karouby, ``{Nonperturbative
  Dynamics Of Reheating After Inflation: A Review},''
  \href{http://dx.doi.org/10.1142/S0218271815300037}{{\em Int. J. Mod. Phys. D}
  {\bfseries 24} (2014) 1530003},
  \href{http://arxiv.org/abs/1410.3808}{{\ttfamily arXiv:1410.3808 [hep-ph]}}.

\bibitem{Giblin:2019nuv}
J.~T. Giblin and A.~J. Tishue, ``{Preheating in Full General Relativity},''
  \href{http://dx.doi.org/10.1103/PhysRevD.100.063543}{{\em Phys. Rev. D}
  {\bfseries 100} no.~6, (2019) 063543},
  \href{http://arxiv.org/abs/1907.10601}{{\ttfamily arXiv:1907.10601 [gr-qc]}}.

\bibitem{Fan:2021otj}
J.~Fan, K.~D. Lozanov, and Q.~Lu, ``{Spillway Preheating},''
  \href{http://dx.doi.org/10.1007/JHEP05(2021)069}{{\em JHEP} {\bfseries 05}
  (2021) 069}, \href{http://arxiv.org/abs/2101.11008}{{\ttfamily
  arXiv:2101.11008 [hep-ph]}}.

\bibitem{Garcia:2021iag}
M.~A.~G. Garcia, K.~Kaneta, Y.~Mambrini, K.~A. Olive, and S.~Verner,
  ``{Freeze-in from preheating},''
  \href{http://dx.doi.org/10.1088/1475-7516/2022/03/016}{{\em JCAP} {\bfseries
  03} no.~03, (2022) 016}, \href{http://arxiv.org/abs/2109.13280}{{\ttfamily
  arXiv:2109.13280 [hep-ph]}}.

\bibitem{Giudice:2000ex}
G.~F. Giudice, E.~W. Kolb, and A.~Riotto, ``{Largest temperature of the
  radiation era and its cosmological implications},''
  \href{http://dx.doi.org/10.1103/PhysRevD.64.023508}{{\em Phys. Rev. D}
  {\bfseries 64} (2001) 023508},
  \href{http://arxiv.org/abs/hep-ph/0005123}{{\ttfamily arXiv:hep-ph/0005123}}.

\bibitem{Chung:1998rq}
D.~J.~H. Chung, E.~W. Kolb, and A.~Riotto, ``{Production of massive particles
  during reheating},'' \href{http://dx.doi.org/10.1103/PhysRevD.60.063504}{{\em
  Phys. Rev. D} {\bfseries 60} (1999) 063504},
  \href{http://arxiv.org/abs/hep-ph/9809453}{{\ttfamily arXiv:hep-ph/9809453}}.

\bibitem{Dudas:2017rpa}
E.~Dudas, Y.~Mambrini, and K.~Olive, ``{Case for an EeV Gravitino},''
  \href{http://dx.doi.org/10.1103/PhysRevLett.119.051801}{{\em Phys. Rev.
  Lett.} {\bfseries 119} no.~5, (2017) 051801},
  \href{http://arxiv.org/abs/1704.03008}{{\ttfamily arXiv:1704.03008
  [hep-ph]}}.

\bibitem{Garcia:2017tuj}
M.~A.~G. Garcia, Y.~Mambrini, K.~A. Olive, and M.~Peloso, ``{Enhancement of the
  Dark Matter Abundance Before Reheating: Applications to Gravitino Dark
  Matter},'' \href{http://dx.doi.org/10.1103/PhysRevD.96.103510}{{\em Phys.
  Rev. D} {\bfseries 96} no.~10, (2017) 103510},
  \href{http://arxiv.org/abs/1709.01549}{{\ttfamily arXiv:1709.01549
  [hep-ph]}}.

\bibitem{Chen:2017kvz}
S.-L. Chen and Z.~Kang, ``{On UltraViolet Freeze-in Dark Matter during
  Reheating},'' \href{http://dx.doi.org/10.1088/1475-7516/2018/05/036}{{\em
  JCAP} {\bfseries 05} (2018) 036},
  \href{http://arxiv.org/abs/1711.02556}{{\ttfamily arXiv:1711.02556
  [hep-ph]}}.

\bibitem{Garcia:2020eof}
M.~A.~G. Garcia, K.~Kaneta, Y.~Mambrini, and K.~A. Olive, ``{Reheating and
  Post-inflationary Production of Dark Matter},''
  \href{http://dx.doi.org/10.1103/PhysRevD.101.123507}{{\em Phys. Rev. D}
  {\bfseries 101} no.~12, (2020) 123507},
  \href{http://arxiv.org/abs/2004.08404}{{\ttfamily arXiv:2004.08404
  [hep-ph]}}.

\bibitem{Garcia:2020wiy}
M.~A.~G. Garcia, K.~Kaneta, Y.~Mambrini, and K.~A. Olive, ``{Inflaton
  Oscillations and Post-Inflationary Reheating},''
  \href{http://dx.doi.org/10.1088/1475-7516/2021/04/012}{{\em JCAP} {\bfseries
  04} (2021) 012}, \href{http://arxiv.org/abs/2012.10756}{{\ttfamily
  arXiv:2012.10756 [hep-ph]}}.

\bibitem{Davidson:2000er}
S.~Davidson and S.~Sarkar, ``{Thermalization after inflation},''
  \href{http://dx.doi.org/10.1088/1126-6708/2000/11/012}{{\em JHEP} {\bfseries
  11} (2000) 012}, \href{http://arxiv.org/abs/hep-ph/0009078}{{\ttfamily
  arXiv:hep-ph/0009078}}.

\bibitem{Harigaya:2013vwa}
K.~Harigaya and K.~Mukaida, ``{Thermalization after/during Reheating},''
  \href{http://dx.doi.org/10.1007/JHEP05(2014)006}{{\em JHEP} {\bfseries 05}
  (2014) 006}, \href{http://arxiv.org/abs/1312.3097}{{\ttfamily arXiv:1312.3097
  [hep-ph]}}.

\bibitem{Harigaya:2014waa}
K.~Harigaya, M.~Kawasaki, K.~Mukaida, and M.~Yamada, ``{Dark Matter Production
  in Late Time Reheating},''
  \href{http://dx.doi.org/10.1103/PhysRevD.89.083532}{{\em Phys. Rev. D}
  {\bfseries 89} no.~8, (2014) 083532},
  \href{http://arxiv.org/abs/1402.2846}{{\ttfamily arXiv:1402.2846 [hep-ph]}}.

\bibitem{Mukaida:2015ria}
K.~Mukaida and M.~Yamada, ``{Thermalization Process after Inflation and
  Effective Potential of Scalar Field},''
  \href{http://dx.doi.org/10.1088/1475-7516/2016/02/003}{{\em JCAP} {\bfseries
  02} (2016) 003}, \href{http://arxiv.org/abs/1506.07661}{{\ttfamily
  arXiv:1506.07661 [hep-ph]}}.

\bibitem{Garcia:2018wtq}
M.~A.~G. Garcia and M.~A. Amin, ``{Prethermalization production of dark
  matter},'' \href{http://dx.doi.org/10.1103/PhysRevD.98.103504}{{\em Phys.
  Rev. D} {\bfseries 98} no.~10, (2018) 103504},
  \href{http://arxiv.org/abs/1806.01865}{{\ttfamily arXiv:1806.01865
  [hep-ph]}}.

\bibitem{Harigaya:2019tzu}
K.~Harigaya, K.~Mukaida, and M.~Yamada, ``{Dark Matter Production during the
  Thermalization Era},'' \href{http://dx.doi.org/10.1007/JHEP07(2019)059}{{\em
  JHEP} {\bfseries 07} (2019) 059},
  \href{http://arxiv.org/abs/1901.11027}{{\ttfamily arXiv:1901.11027
  [hep-ph]}}.

\bibitem{Hall:2009bx}
L.~J. Hall, K.~Jedamzik, J.~March-Russell, and S.~M. West, ``{Freeze-In
  Production of FIMP Dark Matter},''
  \href{http://dx.doi.org/10.1007/JHEP03(2010)080}{{\em JHEP} {\bfseries 03}
  (2010) 080}, \href{http://arxiv.org/abs/0911.1120}{{\ttfamily arXiv:0911.1120
  [hep-ph]}}.

\bibitem{Chu:2011be}
X.~Chu, T.~Hambye, and M.~H.~G. Tytgat, ``{The Four Basic Ways of Creating Dark
  Matter Through a Portal},''
  \href{http://dx.doi.org/10.1088/1475-7516/2012/05/034}{{\em JCAP} {\bfseries
  05} (2012) 034}, \href{http://arxiv.org/abs/1112.0493}{{\ttfamily
  arXiv:1112.0493 [hep-ph]}}.

\bibitem{Mambrini:2013iaa}
Y.~Mambrini, K.~A. Olive, J.~Quevillon, and B.~Zaldivar, ``{Gauge Coupling
  Unification and Nonequilibrium Thermal Dark Matter},''
  \href{http://dx.doi.org/10.1103/PhysRevLett.110.241306}{{\em Phys. Rev.
  Lett.} {\bfseries 110} no.~24, (2013) 241306},
  \href{http://arxiv.org/abs/1302.4438}{{\ttfamily arXiv:1302.4438 [hep-ph]}}.

\bibitem{Chu:2013jja}
X.~Chu, Y.~Mambrini, J.~Quevillon, and B.~Zaldivar, ``{Thermal and non-thermal
  production of dark matter via Z'-portal(s)},''
  \href{http://dx.doi.org/10.1088/1475-7516/2014/01/034}{{\em JCAP} {\bfseries
  01} (2014) 034}, \href{http://arxiv.org/abs/1306.4677}{{\ttfamily
  arXiv:1306.4677 [hep-ph]}}.

\bibitem{Kaneta:2016vkq}
K.~{Kaneta}, Z.~{Kang}, and H.-S. {Lee}, ``{Right-handed neutrino dark matter
  under the B - L gauge interaction},''
  \href{http://dx.doi.org/10.1007/JHEP02(2017)031}{{\em JHEP} {\bfseries 02}
  (2017) 031}, \href{http://arxiv.org/abs/1606.09317}{{\ttfamily
  arXiv:1606.09317 [hep-ph]}}.

\bibitem{Kaneta:2016wvf}
K.~Kaneta, H.-S. Lee, and S.~Yun, ``{Portal Connecting Dark Photons and
  Axions},'' \href{http://dx.doi.org/10.1103/PhysRevLett.118.101802}{{\em Phys.
  Rev. Lett.} {\bfseries 118} no.~10, (2017) 101802},
  \href{http://arxiv.org/abs/1611.01466}{{\ttfamily arXiv:1611.01466
  [hep-ph]}}.

\bibitem{Kaneta:2017wfh}
K.~Kaneta, H.-S. Lee, and S.~Yun, ``{Dark photon relic dark matter production
  through the dark axion portal},''
  \href{http://dx.doi.org/10.1103/PhysRevD.95.115032}{{\em Phys. Rev. D}
  {\bfseries 95} no.~11, (2017) 115032},
  \href{http://arxiv.org/abs/1704.07542}{{\ttfamily arXiv:1704.07542
  [hep-ph]}}.

\bibitem{Bernal:2017kxu}
N.~Bernal, M.~Heikinheimo, T.~Tenkanen, K.~Tuominen, and V.~Vaskonen, ``{The
  Dawn of FIMP Dark Matter: A Review of Models and Constraints},''
  \href{http://dx.doi.org/10.1142/S0217751X1730023X}{{\em Int. J. Mod. Phys. A}
  {\bfseries 32} no.~27, (2017) 1730023},
  \href{http://arxiv.org/abs/1706.07442}{{\ttfamily arXiv:1706.07442
  [hep-ph]}}.

\bibitem{Biswas:2018aib}
A.~Biswas, D.~Borah, and A.~Dasgupta, ``{UV complete framework of freeze-in
  massive particle dark matter},''
  \href{http://dx.doi.org/10.1103/PhysRevD.99.015033}{{\em Phys. Rev. D}
  {\bfseries 99} no.~1, (2019) 015033},
  \href{http://arxiv.org/abs/1805.06903}{{\ttfamily arXiv:1805.06903
  [hep-ph]}}.

\bibitem{Bernal:2019mhf}
N.~Bernal, F.~Elahi, C.~Maldonado, and J.~Unwin, ``{Ultraviolet Freeze-in and
  Non-Standard Cosmologies},''
  \href{http://dx.doi.org/10.1088/1475-7516/2019/11/026}{{\em JCAP} {\bfseries
  11} (2019) 026}, \href{http://arxiv.org/abs/1909.07992}{{\ttfamily
  arXiv:1909.07992 [hep-ph]}}.

\bibitem{Kaneta:2019zgw}
K.~Kaneta, Y.~Mambrini, and K.~A. Olive, ``{Radiative production of nonthermal
  dark matter},'' \href{http://dx.doi.org/10.1103/PhysRevD.99.063508}{{\em
  Phys. Rev. D} {\bfseries 99} no.~6, (2019) 063508},
  \href{http://arxiv.org/abs/1901.04449}{{\ttfamily arXiv:1901.04449
  [hep-ph]}}.

\bibitem{Bernal:2020qyu}
N.~Bernal, J.~Rubio, and H.~Veerm\"ae, ``{UV Freeze-in in Starobinsky
  Inflation},'' \href{http://dx.doi.org/10.1088/1475-7516/2020/10/021}{{\em
  JCAP} {\bfseries 10} (2020) 021},
  \href{http://arxiv.org/abs/2006.02442}{{\ttfamily arXiv:2006.02442
  [hep-ph]}}.

\bibitem{Bernal:2020gzm}
N.~Bernal, ``{Boosting Freeze-in through Thermalization},''
  \href{http://dx.doi.org/10.1088/1475-7516/2020/10/006}{{\em JCAP} {\bfseries
  10} (2020) 006}, \href{http://arxiv.org/abs/2005.08988}{{\ttfamily
  arXiv:2005.08988 [hep-ph]}}.

\bibitem{Anastasopoulos:2020gbu}
P.~Anastasopoulos, K.~Kaneta, Y.~Mambrini, and M.~Pierre, ``{Energy-momentum
  portal to dark matter and emergent gravity},''
  \href{http://dx.doi.org/10.1103/PhysRevD.102.055019}{{\em Phys. Rev. D}
  {\bfseries 102} no.~5, (2020) 055019},
  \href{http://arxiv.org/abs/2007.06534}{{\ttfamily arXiv:2007.06534
  [hep-ph]}}.

\bibitem{Brax:2020gqg}
P.~Brax, K.~Kaneta, Y.~Mambrini, and M.~Pierre, ``{Disformal dark matter},''
  \href{http://dx.doi.org/10.1103/PhysRevD.103.015028}{{\em Phys. Rev. D}
  {\bfseries 103} no.~1, (2021) 015028},
  \href{http://arxiv.org/abs/2011.11647}{{\ttfamily arXiv:2011.11647
  [hep-ph]}}.

\bibitem{Brax:2021gpe}
P.~Brax, K.~Kaneta, Y.~Mambrini, and M.~Pierre, ``{Metastable Conformal Dark
  Matter},'' \href{http://dx.doi.org/10.1103/PhysRevD.103.115016}{{\em Phys.
  Rev. D} {\bfseries 103} no.~11, (2021) 115016},
  \href{http://arxiv.org/abs/2103.02615}{{\ttfamily arXiv:2103.02615
  [hep-ph]}}.

\bibitem{Kaneta:2021pyx}
K.~Kaneta, P.~Ko, and W.-I. Park, ``{Conformal portal to dark matter},''
  \href{http://dx.doi.org/10.1103/PhysRevD.104.075018}{{\em Phys. Rev. D}
  {\bfseries 104} no.~7, (2021) 075018},
  \href{http://arxiv.org/abs/2106.01923}{{\ttfamily arXiv:2106.01923
  [hep-ph]}}.

\bibitem{Ghosh:2022hen}
A.~Ghosh and S.~Mukhopadhyay, ``{Momentum distribution of dark matter produced
  in inflaton decay: effect of inflaton mediated scatterings},''
  \href{http://arxiv.org/abs/2205.03440}{{\ttfamily arXiv:2205.03440
  [hep-ph]}}.

\bibitem{Garny:2015sjg}
M.~Garny, M.~Sandora, and M.~S. Sloth, ``{Planckian Interacting Massive
  Particles as Dark Matter},''
  \href{http://dx.doi.org/10.1103/PhysRevLett.116.101302}{{\em Phys. Rev.
  Lett.} {\bfseries 116} no.~10, (2016) 101302},
  \href{http://arxiv.org/abs/1511.03278}{{\ttfamily arXiv:1511.03278
  [hep-ph]}}.

\bibitem{Garny:2017kha}
M.~Garny, A.~Palessandro, M.~Sandora, and M.~S. Sloth, ``{Theory and
  Phenomenology of Planckian Interacting Massive Particles as Dark Matter},''
  \href{http://dx.doi.org/10.1088/1475-7516/2018/02/027}{{\em JCAP} {\bfseries
  02} (2018) 027}, \href{http://arxiv.org/abs/1709.09688}{{\ttfamily
  arXiv:1709.09688 [hep-ph]}}.

\bibitem{Tang:2017hvq}
Y.~Tang and Y.-L. Wu, ``{On Thermal Gravitational Contribution to Particle
  Production and Dark Matter},''
  \href{http://dx.doi.org/10.1016/j.physletb.2017.10.034}{{\em Phys. Lett. B}
  {\bfseries 774} (2017) 676--681},
  \href{http://arxiv.org/abs/1708.05138}{{\ttfamily arXiv:1708.05138
  [hep-ph]}}.

\bibitem{Chianese:2020yjo}
M.~Chianese, B.~Fu, and S.~F. King, ``{Impact of Higgs portal on
  gravity-mediated production of superheavy dark matter},''
  \href{http://dx.doi.org/10.1088/1475-7516/2020/06/019}{{\em JCAP} {\bfseries
  06} (2020) 019}, \href{http://arxiv.org/abs/2003.07366}{{\ttfamily
  arXiv:2003.07366 [hep-ph]}}.

\bibitem{Chianese:2020khl}
M.~Chianese, B.~Fu, and S.~F. King, ``{Interplay between neutrino and gravity
  portals for FIMP dark matter},''
  \href{http://dx.doi.org/10.1088/1475-7516/2021/01/034}{{\em JCAP} {\bfseries
  01} (2021) 034}, \href{http://arxiv.org/abs/2009.01847}{{\ttfamily
  arXiv:2009.01847 [hep-ph]}}.

\bibitem{Redi:2020ffc}
M.~Redi, A.~Tesi, and H.~Tillim, ``{Gravitational Production of a Conformal
  Dark Sector},'' \href{http://dx.doi.org/10.1007/JHEP05(2021)010}{{\em JHEP}
  {\bfseries 05} (2021) 010}, \href{http://arxiv.org/abs/2011.10565}{{\ttfamily
  arXiv:2011.10565 [hep-ph]}}.

\bibitem{Bernal:2018qlk}
N.~Bernal, M.~Dutra, Y.~Mambrini, K.~Olive, M.~Peloso, and M.~Pierre, ``{Spin-2
  Portal Dark Matter},''
  \href{http://dx.doi.org/10.1103/PhysRevD.97.115020}{{\em Phys. Rev. D}
  {\bfseries 97} no.~11, (2018) 115020},
  \href{http://arxiv.org/abs/1803.01866}{{\ttfamily arXiv:1803.01866
  [hep-ph]}}.

\bibitem{Chung:1998bt}
D.~J.~H. Chung, ``{Classical Inflation Field Induced Creation of Superheavy
  Dark Matter},'' \href{http://dx.doi.org/10.1103/PhysRevD.67.083514}{{\em
  Phys. Rev. D} {\bfseries 67} (2003) 083514},
  \href{http://arxiv.org/abs/hep-ph/9809489}{{\ttfamily arXiv:hep-ph/9809489}}.

\bibitem{Kolb:1998ki}
E.~W. Kolb, D.~J.~H. Chung, and A.~Riotto, ``{WIMPzillas!},''
  \href{http://dx.doi.org/10.1063/1.59655}{{\em AIP Conf. Proc.} {\bfseries
  484} no.~1, (1999) 91--105},
  \href{http://arxiv.org/abs/hep-ph/9810361}{{\ttfamily arXiv:hep-ph/9810361}}.

\bibitem{Chung:2001cb}
D.~J.~H. Chung, P.~Crotty, E.~W. Kolb, and A.~Riotto, ``{On the Gravitational
  Production of Superheavy Dark Matter},''
  \href{http://dx.doi.org/10.1103/PhysRevD.64.043503}{{\em Phys. Rev. D}
  {\bfseries 64} (2001) 043503},
  \href{http://arxiv.org/abs/hep-ph/0104100}{{\ttfamily arXiv:hep-ph/0104100}}.

\bibitem{Ema:2018ucl}
Y.~Ema, K.~Nakayama, and Y.~Tang, ``{Production of Purely Gravitational Dark
  Matter},'' \href{http://dx.doi.org/10.1007/JHEP09(2018)135}{{\em JHEP}
  {\bfseries 09} (2018) 135}, \href{http://arxiv.org/abs/1804.07471}{{\ttfamily
  arXiv:1804.07471 [hep-ph]}}.

\bibitem{Chung:2018ayg}
D.~J.~H. Chung, E.~W. Kolb, and A.~J. Long, ``{Gravitational production of
  super-Hubble-mass particles: an analytic approach},''
  \href{http://dx.doi.org/10.1007/JHEP01(2019)189}{{\em JHEP} {\bfseries 01}
  (2019) 189}, \href{http://arxiv.org/abs/1812.00211}{{\ttfamily
  arXiv:1812.00211 [hep-ph]}}.

\bibitem{Ema:2019yrd}
Y.~Ema, K.~Nakayama, and Y.~Tang, ``{Production of purely gravitational dark
  matter: the case of fermion and vector boson},''
  \href{http://dx.doi.org/10.1007/JHEP07(2019)060}{{\em JHEP} {\bfseries 07}
  (2019) 060}, \href{http://arxiv.org/abs/1903.10973}{{\ttfamily
  arXiv:1903.10973 [hep-ph]}}.

\bibitem{Ahmed:2020fhc}
A.~Ahmed, B.~Grzadkowski, and A.~Socha, ``{Gravitational production of vector
  dark matter},'' \href{http://dx.doi.org/10.1007/JHEP08(2020)059}{{\em JHEP}
  {\bfseries 08} (2020) 059}, \href{http://arxiv.org/abs/2005.01766}{{\ttfamily
  arXiv:2005.01766 [hep-ph]}}.

\bibitem{Gross:2020zam}
C.~Gross, S.~Karamitsos, G.~Landini, and A.~Strumia, ``{Gravitational Vector
  Dark Matter},'' \href{http://dx.doi.org/10.1007/JHEP03(2021)174}{{\em JHEP}
  {\bfseries 03} (2021) 174}, \href{http://arxiv.org/abs/2012.12087}{{\ttfamily
  arXiv:2012.12087 [hep-ph]}}.

\bibitem{Kolb:2020fwh}
E.~W. Kolb and A.~J. Long, ``{Completely dark photons from gravitational
  particle production during the inflationary era},''
  \href{http://dx.doi.org/10.1007/JHEP03(2021)283}{{\em JHEP} {\bfseries 03}
  (2021) 283}, \href{http://arxiv.org/abs/2009.03828}{{\ttfamily
  arXiv:2009.03828 [astro-ph.CO]}}.

\bibitem{Mambrini:2021zpp}
Y.~Mambrini and K.~A. Olive, ``{Gravitational Production of Dark Matter during
  Reheating},'' \href{http://dx.doi.org/10.1103/PhysRevD.103.115009}{{\em Phys.
  Rev. D} {\bfseries 103} no.~11, (2021) 115009},
  \href{http://arxiv.org/abs/2102.06214}{{\ttfamily arXiv:2102.06214
  [hep-ph]}}.

\bibitem{Ling:2021zlj}
S.~Ling and A.~J. Long, ``{Superheavy scalar dark matter from gravitational
  particle production in $\alpha$-attractor models of inflation},''
  \href{http://dx.doi.org/10.1103/PhysRevD.103.103532}{{\em Phys. Rev. D}
  {\bfseries 103} no.~10, (2021) 103532},
  \href{http://arxiv.org/abs/2101.11621}{{\ttfamily arXiv:2101.11621
  [astro-ph.CO]}}.

\bibitem{Haque:2021mab}
M.~R. Haque and D.~Maity, ``{Gravitational dark matter: Free streaming and
  phase space distribution},''
  \href{http://dx.doi.org/10.1103/PhysRevD.106.023506}{{\em Phys. Rev. D}
  {\bfseries 106} no.~2, (2022) 023506},
  \href{http://arxiv.org/abs/2112.14668}{{\ttfamily arXiv:2112.14668
  [hep-ph]}}.

\bibitem{Clery:2021bwz}
S.~Clery, Y.~Mambrini, K.~A. Olive, and S.~Verner, ``{Gravitational portals in
  the early Universe},''
  \href{http://dx.doi.org/10.1103/PhysRevD.105.075005}{{\em Phys. Rev. D}
  {\bfseries 105} no.~7, (2022) 075005},
  \href{http://arxiv.org/abs/2112.15214}{{\ttfamily arXiv:2112.15214
  [hep-ph]}}.

\bibitem{Haque:2022kez}
M.~R. Haque and D.~Maity, ``{Gravitational Reheating},''
  \href{http://arxiv.org/abs/2201.02348}{{\ttfamily arXiv:2201.02348
  [hep-ph]}}.

\bibitem{Clery:2022wib}
S.~Clery, Y.~Mambrini, K.~A. Olive, A.~Shkerin, and S.~Verner, ``{Gravitational
  portals with nonminimal couplings},''
  \href{http://dx.doi.org/10.1103/PhysRevD.105.095042}{{\em Phys. Rev. D}
  {\bfseries 105} no.~9, (2022) 095042},
  \href{http://arxiv.org/abs/2203.02004}{{\ttfamily arXiv:2203.02004
  [hep-ph]}}.

\bibitem{Aoki:2022dzd}
S.~Aoki, H.~M. Lee, A.~G. Menkara, and K.~Yamashita, ``{Reheating and dark
  matter freeze-in in the Higgs-R$^{2}$ inflation model},''
  \href{http://dx.doi.org/10.1007/JHEP05(2022)121}{{\em JHEP} {\bfseries 05}
  (2022) 121}, \href{http://arxiv.org/abs/2202.13063}{{\ttfamily
  arXiv:2202.13063 [hep-ph]}}.

\bibitem{Parker:1969au}
L.~Parker, ``{Quantized fields and particle creation in expanding universes.
  1.},'' \href{http://dx.doi.org/10.1103/PhysRev.183.1057}{{\em Phys. Rev.}
  {\bfseries 183} (1969) 1057--1068}.

\bibitem{Kallosh:2013hoa}
R.~Kallosh and A.~Linde, ``{Universality Class in Conformal Inflation},''
  \href{http://dx.doi.org/10.1088/1475-7516/2013/07/002}{{\em JCAP} {\bfseries
  07} (2013) 002}, \href{http://arxiv.org/abs/1306.5220}{{\ttfamily
  arXiv:1306.5220 [hep-th]}}.

\bibitem{Planck:2018jri}
{\bfseries Planck} Collaboration, Y.~Akrami {\em et~al.}, ``{Planck 2018
  results. X. Constraints on inflation},''
  \href{http://dx.doi.org/10.1051/0004-6361/201833887}{{\em Astron. Astrophys.}
  {\bfseries 641} (2020) A10},
  \href{http://arxiv.org/abs/1807.06211}{{\ttfamily arXiv:1807.06211
  [astro-ph.CO]}}.

\bibitem{Ballesteros:2020adh}
G.~Ballesteros, M.~A.~G. Garcia, and M.~Pierre, ``{How warm are non-thermal
  relics? Lyman-$\alpha$ bounds on out-of-equilibrium dark matter},''
  \href{http://dx.doi.org/10.1088/1475-7516/2021/03/101}{{\em JCAP} {\bfseries
  03} (2021) 101}, \href{http://arxiv.org/abs/2011.13458}{{\ttfamily
  arXiv:2011.13458 [hep-ph]}}.

\bibitem{deGarciaMaia:1993ck}
M.~R. de~Garcia~Maia, ``{Spectrum and energy density of relic gravitons in flat
  Robertson-Walker universes},''
  \href{http://dx.doi.org/10.1103/PhysRevD.48.647}{{\em Phys. Rev. D}
  {\bfseries 48} (1993) 647--662}.

\bibitem{Chernikov:1968zm}
N.~A. Chernikov and E.~A. Tagirov, ``{Quantum theory of scalar fields in de
  Sitter space-time},'' {\em Ann. Inst. H. Poincare Phys. Theor. A} {\bfseries
  9} (1968) 109.

\bibitem{Bunch:1978yq}
T.~S. Bunch and P.~C.~W. Davies, ``{Quantum Field Theory in de Sitter Space:
  Renormalization by Point Splitting},''
  \href{http://dx.doi.org/10.1098/rspa.1978.0060}{{\em Proc. Roy. Soc. Lond. A}
  {\bfseries 360} (1978) 117--134}.

\bibitem{Garcia:2022vwm}
M.~A.~G. Garcia, M.~Pierre, and S.~Verner, ``{Scalar Dark Matter Production
  from Preheating and Structure Formation Constraints},''
  \href{http://arxiv.org/abs/2206.08940}{{\ttfamily arXiv:2206.08940
  [hep-ph]}}.

\end{thebibliography}\endgroup

\end{document}